\documentclass[%
preprint,
 amsmath,amssymb,
physrev 
]{revtex4-2}
\usepackage{amsmath}
\usepackage{graphicx}
\usepackage{dcolumn}
\usepackage{bm}
\usepackage{subcaption}
\usepackage{adjustbox}
\usepackage{amsmath} 
\usepackage{bm}
\usepackage{dutchcal}
\usepackage{tikz}
\usetikzlibrary{positioning,calc}
\usepackage{standalone} 
\usepackage{forest}  
\usepackage{tikz-3dplot}
\usepackage{cellspace}
\usepackage{array} 
\usepackage{multirow}
\usepackage{booktabs}
\usepackage{makecell}

\begin{document}
\preprint{APS/123-QED}

\title{A tensor invariant approach to energy flux in magnetohydrodynamic turbulence}

\author{Conan M. Liptrott}
\affiliation{%
 Centre for Fusion, Space, and Astrophysics, Physics Department, University of Warwick,  Coventry, UK.}%

\author{Sandra C. Chapman}%
\affiliation{Centre for Fusion, Space, and Astrophysics, Physics Department, University of Warwick,  Coventry, UK.}

\author{Bogdan Hnat}
\affiliation{Centre for Fusion, Space, and Astrophysics, Physics Department, University of Warwick,  Coventry, UK.}

\author{Nicholas W. Watkins}
\affiliation{Centre for Fusion, Space, and Astrophysics, Physics Department, University of Warwick,  Coventry, UK.}
\affiliation{Grantham Research Institute on Climate Change and the Environment, London School of Economics and Political Science, London, UK}

\date{\today}

\begin{abstract}
A scale-by-scale analysis of energy flux in the turbulent cascade can be performed using the spatially filtered magnetohydrodynamic (MHD) equations, while the gradient tensor invariants are widely used to characterise the structure of velocity and magnetic fields. 
Physical mechanisms responsible for energy flux require specific field configurations whose strength is quantified by these tensor invariants.
We explore this requirement, showing that the tensor invariants act as proxies for mechanistic energy fluxes under quantifiable conditions. 
As a special case, the purely hydrodynamic contributions to energy flux can be expressed exactly in terms of the invariants of the velocity gradient tensor. 
We also show that the invariants bound the available energy flux for distinct physical mechanisms, formalising the idea that each transfer mechanism requires field configurations with gradients of sufficient strength to support a given energy flux.
Results are illustrated using 3D simulations of freely decaying MHD turbulence. 
\end{abstract}

\maketitle

\section{Introduction}\label{sec:introduction}
\subsection{Overview}

Magnetohydrodynamics (MHD) provides an effective description of many astrophysical plasmas, which often exist in a turbulent state \citep{bavassano82, goldstein1995, biskamp2003, brunoSolarWindTurbulence2013, sahraoui20}.
This turbulence is characterised by nonlinear cascades of ideal invariants across a wide range of scales \citep{alexakis2018}, giving rise to complex behaviour that remains an active area of theoretical research \citep{schekochihin2022}. 

The existence of a turbulent cascade naturally motivates a scale-by-scale analysis of energy flux. A robust construction of energy flux requires a definition of spatial scale which is amenable to theoretical studies and real world data analysis. Spatial filtering \citep{germanoTurbulenceFilteringApproach1992} has been widely used to study hydrodynamic turbulence due to its ability to facilitate such scale-by-scale analyses \citep{germanoTurbulenceFilteringApproach1992,  johnsonEnergyTransferLarge2020, johnson2021, johnson2024}. This approach has since been applied to MHD, enabling detailed studies of energy transfer across scales (e.g.~\cite{kessar2016, aluie2017, offermans2018, bian2019}), with particular emphasis on the application to sub-grid scale modelling of MHD turbulence \citep{alexakis2022}. Extended MHD regimes \citep{manzini2022,benella2026a} and fully kinetic simulations \citep{yangEnergyTransferPressure2017} have also used spatial filtering. A key advantage of this approach is that energy flux can be expressed as field gradients, which in turn can be decomposed into contributions from distinct physical mechanisms as well as into local and non-local contributions to the flux \citep{johnsonEnergyTransferLarge2020, johnson2021, johnson2024, capocciEnergyFluxDecomposition2025}. 
Following \citet{capocciEnergyFluxDecomposition2025}, energy flux in MHD can be partitioned into four subfluxes (Inertial, Maxwell, Dynamo, and Advection), which can in turn be separated by scale locality and decomposed into distinct physical mechanisms, allowing identification of those dominant in driving the turbulence.

Recently, an approach based on invariants of the velocity and magnetic field gradient tensors, and their spatially filtered counterparts, has been used to characterise the spatial structure, or topology, of the fields \citep{chongGeneralClassificationThreedimensional1990, tsinober2001, davidson2004, meneveauLagrangianDynamicsModels2011, dallasStructuresDynamicsSmall2013, consolini2015}.
Part of their appeal is their disambiguation of the ``smallest scales" and the fact that they display a number of universal signatures across a wide variety of turbulent flows \citep{tsinober2001, meneveauLagrangianDynamicsModels2011}, although such universality remains an open question in MHD \citep{dallasStructuresDynamicsSmall2013}.
A further appealing aspect is that multispacecraft measurements allow estimation of field gradients \citep{paschmannAnalysisMethodsMultiSpacecraft, dentonPolynomialReconstructionMagnetic2022}, which can in turn be used to construct the invariants. 
This approach was first applied to \emph{in situ} observations by \citet{consolini2015} using the Cluster Ion Spectrometer (CIS) instrument aboard the Cluster mission, and has subsequently been applied to a wide range of astrophysical systems 
\citep{quattrociocchi2019, bandyopadhyay2020, hnat2021, jiStatisticsGeometricalInvariants2023, zhang2023, hnat2025, quattrociocchi2025, quattrociocchi2026a}.

Taken together, these different approaches raise the natural question of whether particular field configurations, as quantified by the invariants, preferentially contribute to energy transfer, as quantified via the filtering approach \citep{yangEnergyTransferPressure2017, bandyopadhyay2020}. \citet{hnat2025} recently found an ordering of the local energy transfer rate with magnetic field invariants using \emph{in situ} solar wind observations, and \citet{benella2026a} combined the filtering and tensor invariant approaches in simulations of compressible Hall MHD, observing a clear statistical ordering of the turbulent energy flux with the invariants of the velocity gradient tensor.

In this paper, we propose a framework that directly links these invariants to the scale-by-scale transfer of energy in terms of underlying physical mechanisms. 
We show analytically that the energy fluxes are bounded above by functions of the local invariants, with the least upper bound obtained by solving the exact optimisation problem. The second invariants therefore establish an envelope for the energy available for transfer across scales. We further show that, under quantifiable conditions, the strain-rate invariants serve as proxies for the local energy flux, with the sign of the third invariant $\bar{R}_S$ being a reliable predictor of the direction of energy transfer. As a special case, the purely hydrodynamic contributions to energy flux can be expressed exactly in terms of the invariants of the velocity gradient tensor. 
Together these results constitute a semi-empirical framework, which we call the invariant-flux framework, that explicitly connects the tensor invariants to the energy fluxes across scales.

\subsection{Structure and Summary of Results}

The paper is organised as follows.
In Section \ref{sec:background}, we establish the necessary background methodology, including several complementary results that are direct consequences of this methodology and serve as foundational elements for our framework. In Section \ref{sec:invariant_flux_framework}, we present the invariant-flux framework, with Section \ref{sec:flux_bound} deriving the energy flux bounds, and Section \ref{sec:proxy} introducing the energy flux proxy.
Throughout the paper, theoretical results are illustrated using numerical simulations of freely decaying, incompressible 3D MHD turbulence, performed with a pseudospectral code \citep{gomez2005} in a periodic cubic domain of size $[0, 2\pi]^3$. Details of these simulations are provided in Appendix \ref{app:simulations}.
To assist the reader, we provide a concise summary of our main findings in Table \ref{tab:results}. This table presents the energy flux bounds, the invariant flux proxies, and the complementary results. 
For brevity, we present the Inertial and Maxwell terms only. The Advection and Dynamo terms either share the same bounds or have very similar bounds due to symmetries, which are discussed where relevant in the text.

\begin{table}[ht!]
\hspace*{-1.5cm}
\centering
\begin{tabular}{@{} lll @{}}
\toprule
\multicolumn{3}{c}{\shortstack{\textbf{Invariant-Flux Bounds (Section~\ref{sec:flux_bound})}\\[-0pt]\rule{0.45\linewidth}{0.1pt}}} \\  \textbf{Contribution} & \textbf{Local Contribution Bound} & \textbf{Mechanism Bounds (Eq.~\eqref{eq:invariant_bound})} \\
\midrule
Inertial $\Pi^{I, \ell}$ & 

$\begin{aligned}
|\Pi^{I, \ell}_{\mathrm{L}}| \le \begin{cases}
        \frac{2}{\sqrt{3}}  \left(\bar{Q}_\Omega + \bar{Q}_S \right) \left(-\bar{Q}_S \right)^{1/2} & \text{if } \bar{Q}_\Omega \ge 9|\bar{Q}_S| \\
        \frac{2}{9}(\bar{Q}_\Omega - 3\bar{Q}_S)^{3/2} & \text{if } \bar{Q}_\Omega < 9|\bar{Q}_S| 
    \end{cases}
\end{aligned}$
&
 $\begin{aligned}
|\Pi^{I, \ell}_{\mathrm{L},SSS}| 
&\le \ell^2 \frac{2}{\sqrt{3}} \left(-\bar{Q}_S\right)^{3/2} \\
|\Pi^{I, \ell}_{\mathrm{L},\Omega S\Omega}| 
&\le \ell^2 \frac{2}{\sqrt{3}} \left(-\bar{Q}_S\right)^{1/2}\left(\bar{Q}_\Omega\right)
\end{aligned}$
\\ \addlinespace[5pt]
Maxwell $\Pi^{M, \ell}$ & 
$|\Pi^{M, \ell}_{\mathrm{L}}| \le \frac{2}{\sqrt{3}}\ell^2 (-\bar{Q}_S)^{1/2}[ (-\bar{Q}_\Sigma)^2+ 14(-\bar{Q}_\Sigma)\bar{Q}_J + \bar{Q}_J^2 ]^{\frac{1}{2}}$
& 
$\begin{aligned}
|\Pi^{M,\ell}_{\mathrm{L}, S\Sigma\Sigma}| &\le \frac{2}{\sqrt{3}}\ell^2 \left(-\bar{Q}_S\right)^{\frac{1}{2}}\left(-\bar{Q}_\Sigma\right) \\
|\Pi^{M,\ell}_{\mathrm{L}, SJJ}| &\le \frac{2}{\sqrt{3}}\ell^2 \left(-\bar{Q}_S\right)^{\frac{1}{2}}\bar{Q}_J \\
|\Pi^{M,\ell}_{\mathrm{L}, SJ\Sigma}| &\le 4\ell^2 \left[\bar{Q}_J \left(-\bar{Q}_S\right)\left(-\bar{Q}_\Sigma\right)\right]^{1/2}
\end{aligned}$

\\ 
\bottomrule \addlinespace[5pt]
\multicolumn{3}{c}{\shortstack{\textbf{Invariant-Flux Proxies (Section~\ref{sec:proxy})}\\[-0pt]\rule{0.45\linewidth}{0.1pt}}}                           \\
\textbf{Flux}  & \textbf{Proxy} & \textbf{Conditions} \\
\midrule
General $\Pi^{X, \ell}_{\mathrm{L}, SYZ}$
& $\begin{aligned}
    &\text{Sign follows } \bar{R}_S\\
    &\text{Magnitude follows } (\bar{R}_S, \bar{Q}_S) \text{ via Vi\`etes formula} 
\end{aligned}$  
 &$\begin{aligned}
     &\cos(\psi_{2j}) \approx 1 \text{  and any of:}\\ &\text{(i) } \mu_j  \text{ constant} \\
     &\text{(ii) $\langle \mu_j | (\bar{R}_S, \bar{Q}_S) \rangle = \langle \mu_j \rangle$}  \\
     &\text{(iii) $\mu_j = \mu_j(\bar{R}_S, \bar{Q}_S)$}
 \end{aligned}$ \\

$\Pi^{I, \ell}_{\mathrm{L}, SSS}$   & $ 3\ell^2 \bar{R}_S$   & Exact  \\
$\Pi^{I, \ell}_{\mathrm{L}, \Omega S \Omega}$  & $\ell^2(\bar{R}_S - \bar{R}_A)$   & Exact  \\
$\Pi^{I, \ell}_{\mathrm{L}}$ & $\ell^2\left(4\bar{R}_S - \bar{R}_A\right)$ & Exact \\ \addlinespace[5pt]
\toprule
\multicolumn{3}{@{}c@{}}{\shortstack{\textbf{Complementary Results}\\[-0pt]\rule{0.45\linewidth}{0.1pt}}}
\\ \textbf{Use} & \textbf{Formula} & \textbf{Section} \\ \midrule
\makecell[l]{Vi\`etes formula for \\ strain-rate eigenvalues}
& $\lambda_k = 2\sqrt{-\frac{\bar{Q}_S}{3}} \cos\left[\frac{1}{3}\arccos\left(\frac{3\bar{R}_S}{2\bar{Q}_S}\sqrt{-\frac{3}{\bar{Q}_S}}\right)- \frac{2\pi k}{3}\right]$ & (\ref{sec:vietes_formula})
\\
\makecell[l]{Dissipation rates \\ from second invariants}
& $\mathcal{D}_u = -4\nu \bar{Q}_S, \quad 
\mathcal{D}_b= 4\eta \bar{Q}_J$ & (\ref{sec:dissipation_invariants})

 \\ \addlinespace[5pt] 
 \bottomrule
\end{tabular}
\caption{Summary of energy flux bounds and proxies via the coarse-grained tensor invariants. Energy flux quantities $\Pi^{X,\ell}_{Y,Z}$ are defined in Section~\ref{sec:energy_fluxes} and depicted for reference in Fig.~\ref{fig:flux_tree}. The scale-dependent tensor invariants $\bar{Q}_X$ are defined in Section~\ref{sec:invariants}.}
\label{tab:results}
\end{table}

\section{Background Methodology} \label{sec:background}
\subsection{Coarse-Grained MHD Equations}
\noindent We take as our starting point the incompressible magnetohydrodynamic (MHD) equations (e.g.~\cite{priest1982}), 
\begin{eqnarray}
    	\partial_t\bm{u} + \left(\bm{u}\cdot \nabla\right) \bm{u}= \left(\bm{b}\cdot \nabla\right) \bm{b} - \nabla p + \nu \nabla ^{2} \bm{u}, \label{eq:motion2} \\
	\partial_t \bm{b} = \nabla \times \left(\bm{u} \times \bm{b}\right) + \eta \nabla^2 \bm{b}, \label{eq:induction2}\\
	\nabla \cdot \bm{b} = \nabla \cdot \bm{u} = 0, \label{eq:incompressibility2}
\end{eqnarray}
where $\bm{u}(\bm{x},t)$ is the flow velocity, $\bm{b}(\bm{x},t) = \bm{B}(\bm{x},t)/\sqrt{4\pi\rho}$ is the magnetic field in Alfv\'en units, $p$ is sum of thermal and magnetic pressure, $\nu$ is the viscosity, and $\eta$ is the magnetic diffusivity. The Lorentz force, $\bm{j}\times \bm{b} = \left(\bm{b} \cdot \nabla \right) \bm{b} - \nabla \left(b^2/2\right)$, has been written in tension-pressure form. 

To isolate features of a scalar field, $a$, larger than a given scale, $\ell$, we convolve the field with a suitably chosen kernel $G_\ell$ \citep{germanoTurbulenceFilteringApproach1992, johnsonEnergyTransferLarge2020, aluie2017},
\begin{equation}
	\overline{a}_\ell \left(\bm{x}\right) = \int_{\Omega} G_\ell(\bm{r})a \left(\bm{x}+\bm{r}\right) \, d\bm{r}. \label{eq:cg_operation} 
\end{equation}
The resulting \textit{coarse-grained} field $\overline{a}_\ell$ retains features on scales $>\ell$, with scales $<\ell$ attenuated by the convolution. The explicit dependence on scale $\ell$ is hereafter omitted. Applying operation \eqref{eq:cg_operation} to the MHD equations~\eqref{eq:motion2}--\eqref{eq:incompressibility2} gives the spatially coarse-grained MHD equations \citep{aluie2017, alexakis2022},
\begin{eqnarray}
        \partial_t \bar{\bm{u}} + (\bar{\bm{u}} \cdot\nabla )\bar{\bm{u}} =  (\bar{\bm{b}} \cdot\nabla )\bar{\bm{b}}  -\nabla \bar{p} - \nabla \cdot \left(\bm{\tau}^{uu} - \bm{\tau}^{bb}\right) + \nu \nabla^2\bar{\bm{u}}, \label{eq:motion_filtered} \\
    \partial_t \bar{\bm{b}}  + (\bar{\bm{u}} \cdot\nabla )\bar{\bm{b}} - (\bar{\bm{b}} \cdot\nabla )\bar{\bm{u}}= - \nabla \cdot \left(\bm{\tau}^{ub} - \bm{\tau}^{bu}\right)  + \eta\nabla^2\bar{\bm{b}}, \label{eq:induction_filtered} \\
    \nabla \cdot \bar{\bm{b}} = \nabla\cdot \bar{\bm{u}} = 0.\label{eq:incompressible_filtered}
\end{eqnarray}
The second order tensor quantities $\bm{\tau}^{fg}$ are the \textit{subscale stress tensors} given by
\begin{equation}
    \tau^{fg}_{ij}  = \overline{f_ig_j} - \bar{f_i}\bar{g_j}.
\end{equation}
They arise from the filtering of the non-linear terms and describe the force exerted on scales $>\ell$ by the unresolved scales $<\ell$. Evolution equations for the large-scale kinetic energy, $E_\ell^u \equiv \frac{1}{2} |\bar{\bm{u}}|^2$, and large scale magnetic energy, $E_\ell^b \equiv \frac{1}{2} |\bar{\bm{b}}|^2$, are readily obtained from 
equations \eqref{eq:motion_filtered} and \eqref{eq:induction_filtered} by taking the dot product with $\bar{\bm{u}}$ and $\bar{\bm{b}}$, respectively, giving \citep{kessar2016,aluie2017,alexakis2022,manzini2022,capocciEnergyFluxDecomposition2025}
\begin{eqnarray}
    \partial_t E_\ell^u + \nabla \cdot \mathbcal{J}^u = -\Pi^{u,\ell} -\mathcal{W}_\ell- \mathcal{D}_u, \label{eq:large_scale_kinetic}\\
    \partial_t E_\ell^b + \nabla \cdot \mathbcal{J}^b  = -\Pi^{b,\ell} + \mathcal{W}_\ell - \mathcal{D}_b.  \label{eq:large_scale_magnetic}
\end{eqnarray}
The terms involving $ \mathbcal{J}^u$ and $\mathbcal{J}^b$ represent the transport of large-scale kinetic and magnetic energies at scale $\ell$ in space. The term $\mathcal{W}_\ell= \bar{b}_i\bar{b}_j\partial_j\bar{u}_i$ appears with opposite sign in eqs.~\eqref{eq:large_scale_kinetic} and 
\eqref{eq:large_scale_magnetic}, representing an exchange of energy between the flow and magnetic field at scales $>\ell$, and supporting various physical interpretations (see discussion of \citet{aluie2017}). The terms  $\mathcal{D}_u= 2\nu\bar{S}_{ij}\bar{S}_{ij}$ and $\mathcal{D}_b= \eta|\nabla \times \bar{\bm{b}}|^2$ represent the direct dissipation of the large-scale energies. The central quantities of interest in this study are the point-wise energy fluxes
\begin{eqnarray}
        \Pi^{u, \ell} = -\tau_{ij}^{uu} \left(\partial_j \bar{u}_i\right) + \tau_{ij}^{bb} \left(\partial_j \bar{u}_i\right), \label{eq:kinetic_flux}\\
    \Pi^{b, \ell}= -\tau_{ij}^{bu} \left(\partial_j \bar{b}_i\right) + \tau_{ij}^{ub} \left(\partial_j \bar{b}_i\right), \label{eq:magnetic_flux}
\end{eqnarray}
which represent the transfer of kinetic and magnetic energy between scales, defined here to be positive if energy is being transferred from larger to smaller scales. The sum of these, $\Pi^{u, \ell} + \Pi^{b, \ell} = \Pi^{\ell}_{\mathrm{MHD}}$, gives the total energy flux. A scale-by-scale analysis of energy flux can be performed by filtering at different scales (e.g. \cite{capocciEnergyFluxDecomposition2025, manzini2022}). 

\subsection{Energy Flux Decomposition} \label{sec:energy_fluxes}
\noindent The energy fluxes in equations \eqref{eq:kinetic_flux} and \eqref{eq:magnetic_flux} can be decomposed into contributions associated with the terms of the filtered MHD equations~\eqref{eq:motion_filtered}--\eqref{eq:incompressible_filtered} from which they originate. These contributions can be further separated into scale-local and scale-nonlocal components, which can in turn be decomposed according to the physical mechanisms responsible for each energy flux.

\subsubsection{Contribution}
\noindent Following the convention in \citet{capocciEnergyFluxDecomposition2025}, we label the four constituent terms in eqs.~\eqref{eq:kinetic_flux} and \eqref{eq:magnetic_flux} as
\begin{align}
    \Pi^{I, \ell} &= - \tau_{ij}^{uu} \partial_j \bar{u}_i, \label{eq:inertial_flux}\\ 
    \Pi^{M, \ell} &= \tau_{ij}^{bb}  \partial_j \bar{u}_i,\label{eq:maxwell_flux}\\
    \Pi^{A, \ell} &= -\tau_{ij}^{bu}  \partial_j \bar{b}_i, \label{eq:advection_flux}\\
    \Pi^{D, \ell} &= \tau_{ij}^{ub} \partial_j \bar{b}_i, \label{eq:dynamo_flux}
\end{align}
where the superscripts stand for Inertial, Maxwell, Advection, and Dynamo. The sum $\Pi^{I, \ell} + \Pi^{M, \ell} = \Pi^{u, \ell}$ gives the flux of large-scale kinetic energy and $\Pi^{A, \ell} + \Pi^{D, \ell} = \Pi^{b, \ell}$ gives the flux of large-scale magnetic energy. 

\subsubsection{Locality}
\noindent It was demonstrated by \citet{johnsonEnergyTransferLarge2020}, and extended to MHD in \citet{capocciEnergyFluxDecomposition2025}, that specifying a Gaussian low-pass filter, 
\begin{equation}
	G_\ell(\mathbf{r}) = \frac{1}{(2\pi\ell^2)^{3/2}}\exp \left(-\frac{|\mathbf{r}|^2}{2\ell^2}\right), \label{eq:gaussian_filter}
\end{equation}
allows for an analytic expression of the subscale stresses $\tau^{fg}_{ij}$ in terms of field gradients $\bar{A}_{ij} = \partial_j \bar{u}_i$ and $\bar{B}_{ij} = \partial_j \bar{b}_i$. This approach naturally separates flux contributions into \textit{scale local} and \textit{scale non-local} components. For example, the Inertial and Maxwell fluxes become
\begin{align}
	\Pi^{I, \ell} &= -\ell^2 \bar{A}_{ij}\bar{A}_{ik}\bar{A}_{jk} - \bar{A}_{ij}  \int_0^{\ell^2} \tau^{\beta}(\bar{A}_{ik}^{ \sqrt{ \alpha }}, \bar{A}_{ik}^{ \sqrt{ \alpha }})  d \alpha \equiv \Pi^{I, \ell}_{\mathrm{L}} + \Pi^{I, \ell}_{\mathrm{NL}},  \label{eq:inertial_local_nonlocal}\\
	\Pi^{M, \ell} &= \ell^2 \bar{A}_{ij}\bar{B}_{ik}\bar{B}_{jk} + \bar{A}_{ij}  \int_0^{\ell^2} \tau^{\beta}(\bar{B}_{ik}^{ \sqrt{ \alpha }}, \bar{B}_{ik}^{ \sqrt{ \alpha }})  d \alpha \equiv \Pi^{M, \ell}_{\mathrm{L}} + \Pi^{M, \ell}_{\mathrm{NL}}. \label{eq:maxwell_local_nonlocal}
\end{align}
Here the integrand subscale terms $\tau^\beta(x,y) = \overline{xy}^\beta - \bar{x}^\beta\bar{y}^\beta$ are filtered at scale $\beta = \sqrt{\ell^2-\alpha}$. The first terms involve the gradients filtered at scale $\ell$ only, representing the local transfer of energy to immediately adjacent scales. The integral terms represent cumulative contributions from all smaller scales, constituting the non-local flux. Similar expressions exist for the Advection and Dynamo terms. 

\subsubsection{Mechanism}

\noindent The physical mechanisms driving these fluxes can be revealed by substitution of a suitable decomposition of the velocity and magnetic field gradient tensors \citep{johnsonEnergyTransferLarge2020, capocciEnergyFluxDecomposition2025}. The coarse-grained gradient tensors can be decomposed into their symmetric and antisymmetric parts. The coarse-grained velocity gradient tensor becomes $\bar{A}_{ij} =\bar{S}_{ij} + \bar{\Omega}_{ij}$, where $\bar{S}_{ij}$ is the flow strain-rate tensor and $\bar{\Omega}_{ij}$ is the  rotation-rate tensor. Similarly, the coarse-grained magnetic field gradient tensor becomes $\bar{B}_{ij} =\bar{\Sigma}_{ij} + \bar{J}_{ij}$, where $\bar{\Sigma}_{ij}$ is the magnetic strain-rate tensor and $\bar{J}_{ij}$ is the current density tensor. The rotation-rate tensor is the dual of the vorticity, $\bar{\Omega}_{ij} = -\frac{1}{2}\epsilon_{ijk}\bar{\omega}_k$, and the current density tensor is the dual of the current density, $\bar{J}_{ij} = -\frac{1}{2}\epsilon_{ijk}\bar{j}_k$. 
Substituting these decompositions into the local flux expressions yields terms that can be interpreted as physical mechanisms. The local terms $\Pi^{I,\ell}_{\mathrm{L}}$ and $\Pi^{M,\ell}_{\mathrm{L}}$ become
\begin{gather}
    \Pi^{I, \ell}_{\mathrm{L}} = -\ell^2 \bar{S}_{ij}\bar{S}_{ik}\bar{S}_{jk} + \frac{\ell^2}{4} \bar{ \omega}_i \bar{S}_{ij} \bar{ \omega}_j, \label{eq:inertial_decomposition} \\
    \Pi^{M,\ell}_{\mathrm{L}} = \ell^2 \bar{S}_{ij} \bar{\Sigma}_{ik} \bar{\Sigma}_{jk} + \ell^2 \bar{S}_{ij} \bar{J}_{ik} \bar{J}_{jk} +2\ell^2 \bar{S}_{ij} \bar{J}_{ik} \bar{\Sigma}_{jk}. \label{eq:maxwell_decomposition}
\end{gather}
In Eq.~\eqref{eq:inertial_decomposition}, the $S_{ij}S_{ik}S_{jk}$ term is the contribution to the kinetic energy flux from strain self-amplification, and the $\bar{\omega}_i S_{ij} \bar{\omega}_j$ term  is the contribution from vortex stretching. These well-established mechanisms from hydrodynamic turbulence are generalised to MHD by the Maxwell, Advection, and Dynamo terms. For example, the $\bar{S}_{ij}\bar{J}_{ik}\bar{J}_{jk}$ term in Eq.~\eqref{eq:maxwell_decomposition} represents the stretching of current filaments in analogy to the vortex stretching mechanism, and the $\bar{S}_{ij}\bar{\Sigma}_{ik}\bar{\Sigma}_{jk}$ term represents magnetic shear amplification, both of which are related to straining motions of the flow (see \cite{capocciEnergyFluxDecomposition2025} for more details). 
    
\begin{figure}[ht!]
    \centering
    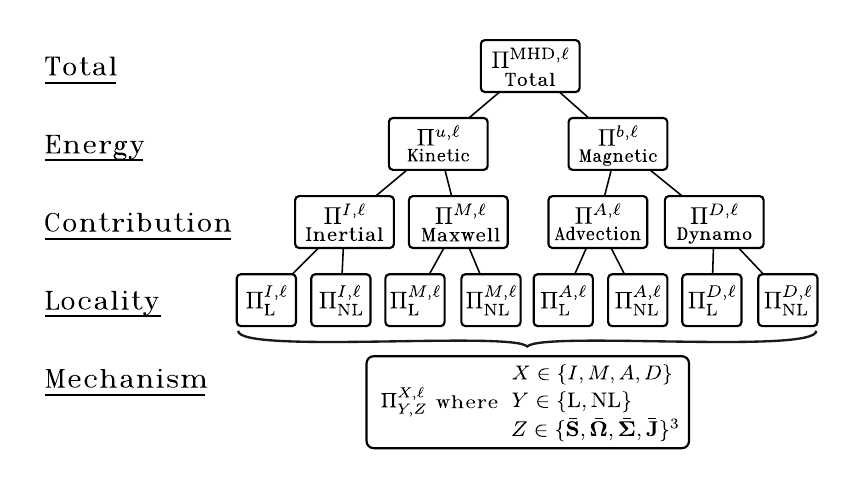
    \caption{Hierarchical decomposition of the total MHD energy flux showing the notation used throughout the paper.}
    \label{fig:flux_tree}
\end{figure}

A fully decomposed flux term takes the general form $\Pi^{X,\ell}_{Y, Z}$ where $X \in \{I,M,A,D\}$ specifies the contribution term, $Y\in \{\mathrm{L}, \mathrm{NL}\}$ specifies the locality, and $Z\in \{\bar{\bm{S}}, \bar{\bm{\Omega}}, \bar{\bm{\Sigma}}, \bar{\bm{J}}\}^3$ denotes the physical mechanism. For example, the local strain self-amplification term in Eq.~\eqref{eq:inertial_decomposition} is denoted $\Pi^{I, \ell}_{\mathrm{L}, SSS}$. A full list of all non-zero terms can be found in \citet{capocciEnergyFluxDecomposition2025}. This hierarchical decomposition of energy flux is visualised in Fig.~\ref{fig:flux_tree} for ease of reference.

\subsection{Energy Flux in Tensor-Angle Form} \label{sec:flux_tensor_angle_form}
 
\noindent The energy fluxes can be expressed in terms of the alignment between the eigenvectors of the tensors \citep{ballouz2020, johnson2021, capocciEnergyFluxDecomposition2025}. This is central to the invariant-flux framework. To illustrate the idea consider the current filament stretching term $\Pi^{M, \ell}_{\mathrm{L}, SJJ}$. Writing the eigenvalues of the strain-rate tensor as $\lambda_i$, the flux is given by
\begin{equation}
    \Pi^{M, \ell}_{\mathrm{L}, SJJ} = -\frac{1}{4} \ell^2|\bar{\bm{j}}|^2\sum_i\lambda_i\cos^2(\theta_{i}), \label{eq:sjj_angle_form}
\end{equation}
where $\theta_{i}$ is the angle between $\bar{\bm{j}}$ and the i-th eigenvector of $\bm{S}$. Details are given in Appendix \ref{app:vector-angle_form}. The energy flux is thus determined by three factors: (i) the magnitude of the current density; (ii) the eigenvalues $\lambda_i$ of the strain-rate tensor; and (iii) the geometric alignment $\cos(\theta_{i})$ between the current density and the eigenvectors of the strain-rate tensor. Perfect misalignment of the current density with the strain-rate eigenvectors will result in minimal energy flux for this term, regardless of the magnitude of current density or strain. A preferential alignment with a single eigenvector, $\hat{\bm{\lambda}}^{(i)}$, so that $\cos^2(\theta_{i}) \approx 1$ and $\cos^2(\theta_{j}) \approx 0$ for $j \ne i$, will lead to a flux dominated in sign and proportional in magnitude to eigenvalue $\lambda_i$. 
A general local flux term can be written 
\begin{equation}
    \Pi^{X,\ell}_{\mathrm{L}, SYZ} \propto \bar{S}_{ij} \bar{Y}_{ik} \bar{Z}_{jk} = \sum_{i,j}\lambda_i \mu_j \cos^2(\psi_{ij}), \label{eq:flux_angle_general_form}
\end{equation}
where $\mu_j$ is the j-th eigenvalue of the symmetric part of the product matrix $\bar{Y}_{ik} \bar{Z}_{kj}$, and $\psi_{ij}$ is the angle between the i-th eigenvector of $\bar{\bm{S}}$ and the j-th eigenvector of the symmetric part of $Y_{ik}^\ell Z_{kj}^\ell$ \citep{capocciEnergyFluxDecomposition2025}. In the case of preferential alignment, $\cos{\psi_{ij}} \approx 1$, the eigenvalue product $\mu_i$ and $\lambda_j$ will determine the sign and be proportional to the magnitude of energy flux. In Appendix \ref{app:tensor_angle_vector_angle} we show that the general case in Eq.~\eqref{eq:flux_angle_general_form} collapses to the vector-angle form in Eq.~\eqref{eq:sjj_angle_form} for $\Pi^{M, \ell}_{\mathrm{L}, SJJ}$.

\subsection{Invariants of the Coarse-Grained Gradient Tensors} \label{sec:invariants}
\noindent This section introduces the tensor invariants of the coarse-grained fields, which can be used to characterise the structure of the fields. Some useful results are also presented that will aid the analysis of the energy fluxes.

\subsubsection{Tensor Invariants}
\noindent A second order tensor $\bm{X}$ can be characterised by three principal invariants arising as the coefficients of its characteristic equation,
\begin{equation}
    \mathrm{det}(\bm{X}-\lambda \bm{I}) = \lambda^3 + P\lambda^2 + Q\lambda + R = 0,\label{eq:characteristic_equation}
\end{equation}
whose solutions are the eigenvalues $\lambda_i$ of $\bm{X}$. As functions of the eigenvalues, the principal invariants do not change under coordinate transformations. The invariants take the forms  
\begin{align}
    P &= -\mathrm{tr}\left(\bm{X}\right) = -(\lambda_1 + \lambda_2 + \lambda_3), \label{eq:P_invariant}\\
    Q  &= \frac{1}{2}\left[P^2 -\mathrm{tr}\left(\bm{X}^2\right)\right] = \lambda_1\lambda_2 + \lambda_2\lambda_3 + \lambda_3\lambda_1,\label{eq:Q_invariant} \\
    R  &= -\mathrm{det}\left(\bm{X}\right) = -\lambda_1 \lambda_2\lambda_3,\label{eq:R_invariant}
\end{align}
where the trace/determinant forms are found by application of the Cayley-Hamilton theorem. If $\bm{X} = \partial_j x_i$ is the gradient of a vector field, its trace, $\mathrm{tr}(\bm{X}) = \partial_i x_i$, is the divergence and hence zero for the divergence-free fields considered here. 
It follows that $P=0$, and the field is characterised by the second and third invariants, $Q$ and $R$, respectively. 
These invariants, sometimes called the topological or geometric invariants, are much used in the hydrodynamics literature \citep{meneveauLagrangianDynamicsModels2011, davidson2004, chongGeneralClassificationThreedimensional1990, tsinober2001} and more recently in the MHD literature \citep{dallasStructuresDynamicsSmall2013,consolini2015,quattrociocchi2019,hnat2021,jiStatisticsGeometricalInvariants2023,zhang2023,hnat2025, quattrociocchi2026a}
 to characterise the local spatial geometry of the flow and magnetic field. 
The invariants of the \textit{coarse-grained} velocity and magnetic field gradient tensors used in this work are stated below. We refer to \citet{dallasStructuresDynamicsSmall2013} for a more complete discussion in the MHD context. The constituent strain-rate and rotation tensors possess their own sets of invariants that can refine the analysis.

\paragraph{Velocity Gradient Invariants}
The invariants of the coarse-grained velocity gradient tensor, $\bar{A}_{ij} = \bar{S}_{ij} + \bar{\Omega}_{ij}$, are
\begin{gather*}
    \bar{Q}_A = \frac{1}{4} \left( \bar{\omega}^2 - 2 \mathrm{tr}(\bar{\bm{S}}^2) \right),\\
    \bar{R}_A = -\frac{1}{3} \left( \mathrm{tr}(\bar{\bm{S}}^3) + \frac{3}{4} \bar{\omega}_i \bar{S}_{ij} \bar{\omega}_j \right).
\end{gather*}
The invariants of the flow strain-rate tensor $\bar{S}_{ij}$ are
\begin{gather*}
    \bar{Q}_{S} = -\frac{1}{2} \mathrm{tr}(\bar{\bm{S}}^2) = -\frac{1}{2} \bar{S}_{ij} \bar{S}_{ij}, \\
    \bar{R}_{S} = -\frac{1}{3} \mathrm{tr}(\bar{\bm{S}}^3) = -\frac{1}{3} \bar{S}_{ij} \bar{S}_{jk} \bar{S}_{ki}.
\end{gather*}
The invariants of the rotation rate tensor $\bar{\Omega}_{ij}$ are 
\begin{gather*}
    \bar{Q}_{\Omega} = -\frac{1}{2} \mathrm{tr}(\bar{\bm{\Omega}}^2) = \frac{1}{4} \bar{\omega}^2 = \bar{Q}_A - \bar{Q}_S, \\
    \bar{R}_{\Omega} = 0.
\end{gather*}
Although the second invariants are additive, $\bar{Q}_A = \bar{Q}_S + \bar{Q}_{\Omega}$, the third invariants are not, $\bar{R}_A \neq \bar{R}_S + \bar{R}_{\Omega} = \bar{R}_S $. It can be seen that the second invariant $\bar{Q}_A$ measures the relative magnitude of vorticity to strain. If $\bar{Q}_A\gg0$, the filtered flow is dominated by rotation. For this reason $\bar{Q}_A$ has been widely used as a vortex identification criterion, the Q-criterion, with subsequent related criteria (e.g. the $\Delta$-criterion \citep{chongGeneralClassificationThreedimensional1990}). 

\paragraph{Magnetic Field Gradient}
Similarly, the invariants of the coarse-grained magnetic fields gradient tensor, $\bar{B}_{ij} = \bar{\Sigma}_{ij} + \bar{J}_{ij}$, are given by
\begin{gather*}
    \bar{Q}_B = \frac{1}{4} \left( \bar{j}^2 - 2 \mathrm{tr}(\bar{\bm{\Sigma}}^2) \right), \\
    \bar{R}_B = -\frac{1}{3} \left( \mathrm{tr}(\bar{\bm{\Sigma}}^3) + \frac{3}{4} \bar{j}_i \bar{\Sigma}_{ij} \bar{j}_j \right).
\end{gather*}
The invariants for the magnetic strain-rate tensor $\bar{\Sigma}_{ij}$ are
\begin{gather*}
    \bar{Q}_{\Sigma} = -\frac{1}{2} \mathrm{tr}(\bar{\bm{\Sigma}}^2) = -\frac{1}{2} \bar{\Sigma}_{ij} \bar{\Sigma}_{ij}, \\
    \bar{R}_{\Sigma} = -\frac{1}{3} \mathrm{tr}(\bar{\bm{\Sigma}}^3) = -\frac{1}{3} \bar{\Sigma}_{ij} \bar{\Sigma}_{jk} \bar{\Sigma}_{ki}.
\end{gather*}
The invariants of the current density tensor $\bar{J}_{ij}$ are
\begin{gather*}
    \bar{Q}_{J} = -\frac{1}{2} \mathrm{tr}(\bar{\bm{J}}^2) = \frac{1}{4} \bar{j}^2 = \bar{Q}_B - \bar{Q}_\Sigma,\\
    \bar{R}_J = 0.
\end{gather*}

\subsubsection{Dissipation Rates From Second Invariants} \label{sec:dissipation_invariants}

\noindent The dissipation rates of large-scale kinetic and magnetic energies, $\mathcal{D}_u$ and $\mathcal{D}_b$, can be written directly in terms of the second invariants of the flow strain-rate tensor and current density tensor, respectively, as
\begin{eqnarray}
    \bar{Q}_S = -\frac{1}{4\nu}\mathcal{D}_u, \quad \bar{Q}_J = \frac{1}{4\eta}\mathcal{D}_b. \label{eq:invariant_dissipation}
\end{eqnarray}
Thus $\bar{Q}_s$ and $\bar{Q}_J$ are measures of the dissipation rates of large-scale kinetic and magnetic energy in the same way as the non coarse-grained counterparts are related to viscous and ohmic dissipation through $Q_S = -\frac{1}{4}\epsilon_u/\nu$ and $Q_J = \frac{1}{4}j^2/\eta$ \citep{dallasStructuresDynamicsSmall2013}. Note that $\bar{Q}_S\le0$ but $\bar{Q}_J\ge0$, accounting for the difference in sign. 

\subsubsection{Strain-Rate Eigenvalues and Third Invariant} \label{sec:strain_rate_third_invariant}

\noindent The properties of the strain-rate eigenvalues reveal the physical meaning of the third invariant, and are central to the later development of the invariant-flux framework. Since the flow strain-rate tensor $\bar{\bm{S}}$ is real, symmetric, and trace-free, it has three real eigenvalues $\lambda_i$ that must satisfy $\lambda_1 + \lambda_2 + \lambda_3 = 0$. Ordering these as $\lambda_1 \ge \lambda_2 \ge \lambda_3$, the non-trivial case requires $\lambda_1>0$ and $\lambda_3<0$, with $\lambda_2$ enforcing the trace-free condition. The eigenvalues are the three principal rates of strain which can either be extensional, $\lambda_i>0$, or compressive, $\lambda_i<0$. If $\lambda_2 > 0$ then we have two extensional directions and one compressive. A spherical blob of fluid will deform into a ``sheet-like'' or ``disk-like'' structure under this straining. If $\lambda_2 < 0$, we have two compressive directions and one extensional, and a spherical blob will deform into a ``tube-like'' structure.
With the sign of $\lambda_{1,3}$ fixed, the sign of the third invariant $\bar{R}_S = -\lambda_1 \lambda_2 \lambda_3$ takes the sign of the intermediate eigenvalue $\lambda_2$. Thus $\bar{R}_S > 0$ corresponds to “sheet-like” structures and $\bar{R}_S < 0$ corresponds to “tube-like” structures \citep{dallasStructuresDynamicsSmall2013}. 

\subsubsection{Vi\`etes Formula for the Eigenvalues} \label{sec:vietes_formula}
\noindent A useful property of the strain-rate eigenvalues will help establish the expected magnitude of the energy fluxes in $(\bar{R}_S, \bar{Q}_S)$ space. 
The strain-rate tensor $\bar{S}_{ij}$ has three real eigenvalues, so Vi\`ete's formula allows us to write these eigenvalues in terms of the invariants as 
\begin{eqnarray}
    \lambda_k = 2\sqrt{-\frac{\bar{Q}_S}{3}} \cos\left[\frac{1}{3}\arccos\left(\frac{3\bar{R}_S}{2\bar{Q}_S}\sqrt{-\frac{3}{\bar{Q}_S}}\right) - \frac{2\pi k}{3}\right], \quad \mathrm{for}\ k = 0,1,2. \label{eq:vietes_formula}
\end{eqnarray}
Three properties are clear from this formulation: (i) the eigenvalue magnitude scales with $\sqrt{-\bar{Q}_S}$; (ii) for a given $\bar{Q}_S$, the eigenvalue attains its maximum and minimum when the argument of the arccosine is $\pm1$, i.e. when $\bar{R}_S = \mp\frac{2}{3\sqrt{3}}(-\bar{Q}_S)^{3/2}$; and (iii) the eigenvalue is zero when the argument of the arccosine is zero, i.e. when $\bar{R}_S = 0$. The extrema occur along $\Delta = 0$, where $\Delta$ the discriminant of the strain-rate tensor, giving the maximum absolute value that $\bar{R}_S$ can take for a given $\bar{Q}_S$. From Eq.~\eqref{eq:invariant_dissipation}, scaling with $\sqrt{-\bar{Q}_S}$ is equivalent to scaling with $\sqrt{\mathcal{D}_u/\nu}$. 
We therefore expect dissipation to bound the energy flux in a way made precise in Section \ref{sec:flux_bound}. Finally, only the intermediate eigenvalue can vanish in the non-trivial case. The remaining two eigenvalues are equal and opposite, taking the values $\lambda_{1}, \lambda_3 = \pm\sqrt{-\bar{Q}_S}$. A mathematically identical formulation applies to the magnetic strain-rate eigenvalues. 

\section{Invariant-Flux Framework} \label{sec:invariant_flux_framework}

\noindent We now propose a framework for understanding and predicting energy flux, down to the level of physical mechanism, from the tensor invariants. We show that the invariants locally bound the magnitude of energy flux (Section \ref{sec:flux_bound}), and detail the conditions under which the invariants act as proxies for the energy flux (Section \ref{sec:proxy}). 
As a special case, the hydrodynamic flux terms can be written exactly in terms of the invariants (Section \ref{sec:hydrodynamic_case}). 

\subsection{Energy Flux Bounds} \label{sec:flux_bound}

\noindent Each energy flux term represents a specific physical mechanism that requires a particular field configuration to operate. Vortex stretching, for instance, requires both vorticity and strain. The strength of each such configuration, given by the magnitude of the gradients, is quantified by the tensor invariants. It follows that the invariants will bound the energy fluxes. 
The method of bounding the flux given known invariants is demonstrated on the simplest case, the strain self-amplification flux in eigenvalue form, 
\begin{equation*}
    \Pi^{I, \ell}_{\mathrm{L}, SSS} = -\ell^2 \mathrm{tr}(\bar{\bm{S}}^3) = -\ell^2\sum_i\lambda_i^3. 
\end{equation*}
We seek the maximum of the eigenvalue sum, $\sum_i\lambda^3_i$, subject to the incompressibility constraint, $\sum_i\lambda_i = 0$, and the known invariant, $\bar{Q}_S = -\frac{1}{2}\sum_i\lambda_i^2$. This is readily achieved by constructing the Lagrange multiplier,
\begin{equation}
\mathcal{L} = \sum_i\lambda_i^3 - \alpha \sum_i\lambda_i - \beta \left(\frac{1}{2}\sum_i\lambda_i^2 + \bar{Q}_S\right). \notag \label{eq:sss_lagrangian}
\end{equation}
Taking derivatives to find the maxima gives 
\begin{align*}
     \mathcal{L}_{\lambda_i} &= 3\lambda_i^2 - \alpha  + \beta \lambda_i = 0. \\
     \mathcal{L}_{\alpha} &= \sum_i \lambda_i = 0,\\
     \mathcal{L}_{\beta} &= -\frac{1}{2}\sum_i\lambda_i^2 - \bar{Q}_S = 0.
\end{align*}
These can be solved simultaneously but it is easier to recognise that all of the eigenvalues must satisfy the same quadratic, which can have at most two distinct roots. Let these roots be $a$ and $b$, and then take $\lambda_1 = \lambda_2 = a$ and $\lambda_3=b$ without loss of generality. Substituting into the  constraints yields
\begin{align*}
    2a + b &= 0,\\
    2a^2 + b^2 &= -2\bar{Q}_S, 
\end{align*}
so that $b = -2a$ and then $a = \pm\sqrt{\frac{-\bar{Q}_S}{3}}$, $b = \mp2\sqrt{\frac{-\bar{Q}_S}{3}}$. Substituting these back into the objective $\sum_i \lambda_i^3$ gives a maximum of 
\begin{equation*}
    2\left(-\sqrt{\frac{-\bar{Q}_S}{3}}\right)^3 + \left(2\sqrt{\frac{-\bar{Q}_S}{3}}\right)^3 = \frac{2}{\sqrt{3}}\left(-\bar{Q}_S\right)^{3/2}.  
\end{equation*}
Reintroducing the scale dependency we obtain the bound on the energy flux 
\begin{equation} 
    |\Pi^{I, \ell}_{\mathrm{L}, SSS}| \le   \ell^2\frac{2}{\sqrt{3}}\left(-\bar{Q}_S\right)^{3/2}.\label{eq:sss_bound}
\end{equation}
In general, a local mechanistic flux term can be bounded above by
\begin{equation}
     |\Pi_{XYZ}|
     \le K |\ell^2| \sqrt{ (\mathcal{s}_X Q_X) (\mathcal{s}_Y Q_Y) (\mathcal{s}_Z Q_Z)}, \quad \mathcal{s}_X = \begin{cases}
    -1 & \bar{\bm{X}}\ \ \mathrm{symmetric}, \\
    1 & \bar{\bm{X}} \ \  \mathrm{antisymmetric},
    \end{cases} \label{eq:invariant_bound}
\end{equation}
where the constant $K$ takes values listed in Table~\ref{tab:results} for each term and is found using the same Lagrange multiplier approach as above.
\emph{The second invariants of the coarse-grained gradient tensors therefore act as an envelope for the possible energy flux.} This is our first main result. Given values of the tensor invariants of the velocity and magnetic field gradient tensors, we can say with certainty that the contribution to energy flux will not exceed the corresponding bound. 

\begin{figure}[ht!]
     \centering
     \begin{subfigure}[b]{0.48\textwidth}
        \centering
    \includegraphics[width=1\linewidth]{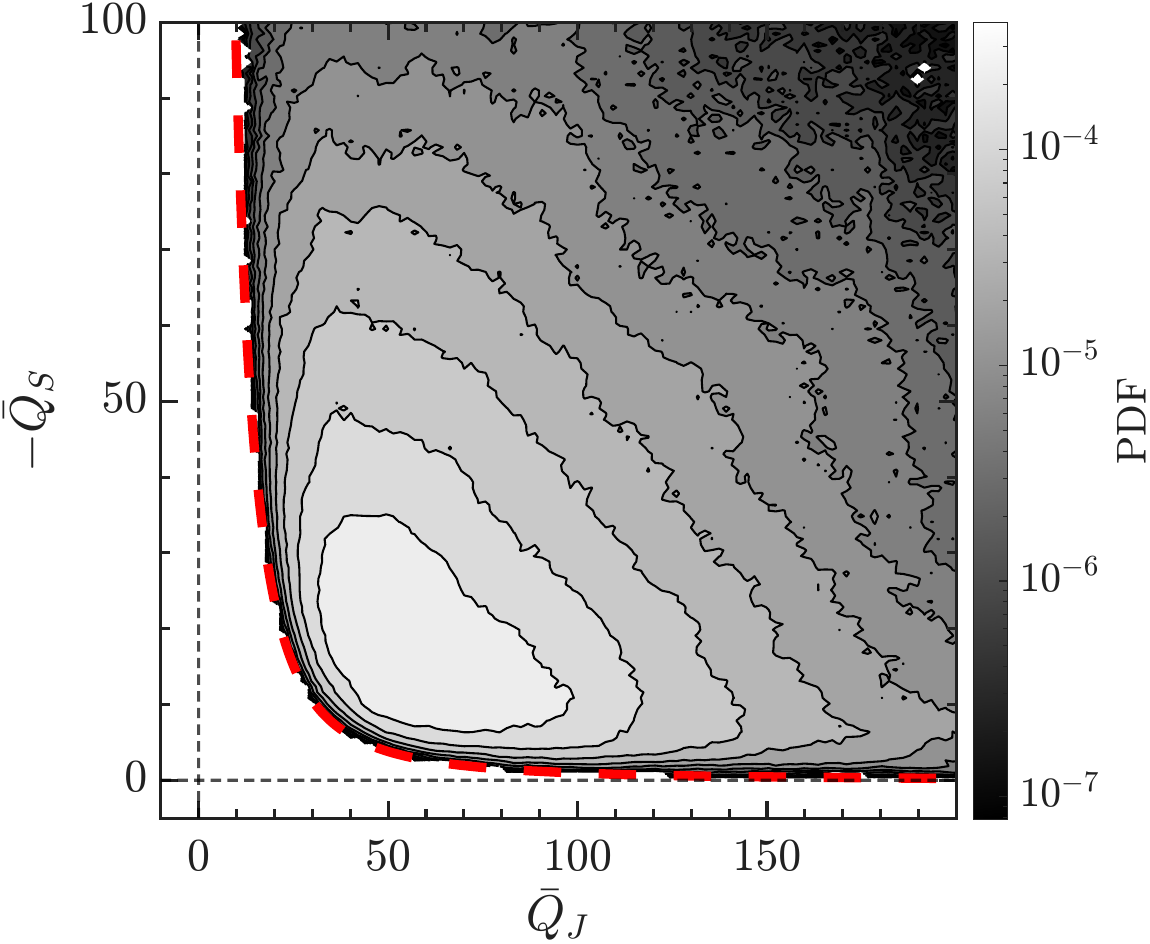}
        \caption{}
        \label{fig:sjj_bound}
     \end{subfigure}
     \begin{subfigure}[b]{0.48\textwidth}
        \centering
    \includegraphics[width=1\linewidth]{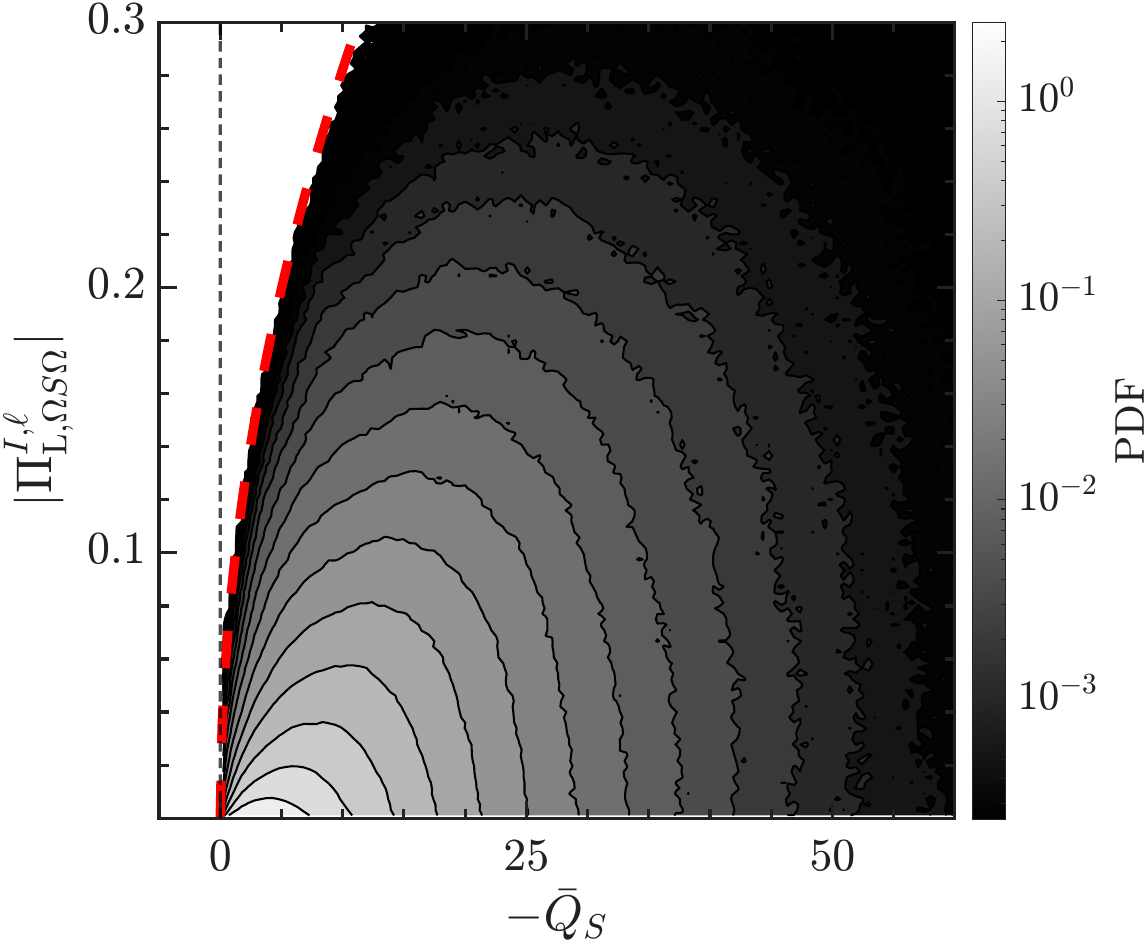}
        \caption{}
        \label{fig:oso_bound}
     \end{subfigure}
     
     \caption{(a) Joint PDF of $(\bar{Q}_J, -\bar{Q}_S)$ for data filtered by $|\Pi^{M, \ell}_{\mathrm{L}, SJJ}|\ge0.5$, with the red dashed line corresponding to bound Eq.~\eqref{eq:sjj_bound}. (b) Joint PDF of $(|\Pi^{I, \ell}_{\mathrm{L}, \Omega S \Omega}|, -\bar{Q}_S)$ for data filtered by $\bar{Q}_\Omega < \langle \bar{Q}_\Omega\rangle$, with the red dashed line corresponding to bound Eq.~\eqref{eq:oso_bound}. Obtained using numerical simulations of freely decaying, incompressible 3D MHD turbulence, performed with a pseudospectral code \citep{gomez2005} in a periodic cubic domain of size $[0, 2\pi]^3$ filtered at scale $\ell\approx47\eta$. Simulation details in Appendix~\ref{app:simulations}. }
     \label{fig:simple_bounds}
\end{figure}

These bounds can be visualised from the simulation data by showing either how the invariants are constrained by a fixed flux value, or how the flux is bounded by fixed invariant values. To visualise how a specific flux level restricts the allowed invariant values, we filter for high-activity regions where $|\Pi|\ge\Pi_{\mathrm{min}}$. Since the bound gives the maximum flux achievable for a given set of invariants, any point in the filtered data must possess invariants large enough to support at least $\Pi_{\mathrm{min}}$. Points whose invariants are insufficient to reach $\Pi_{\mathrm{min}}$ are therefore absent from the filtered data, leaving a cleared region demarcated by the bound evaluated at $\Pi_{\mathrm{min}}$. This is shown for the $\Pi^{M, \ell}_{\mathrm{L}, SJJ}$ term in Fig.~\ref{fig:sjj_bound}, where the data is filtered at $|\Pi^{M,\ell}_{\mathrm{L}, SJJ}|\ge0.5$ and the red dashed line shows the bound evaluated at this threshold. From Eq.~\eqref{eq:invariant_bound} and Table~\ref{tab:results}, this bound is given by 
\begin{equation}
    |\Pi^{M,\ell}_{\mathrm{L}, SJJ}| \le \frac{2}{\sqrt{3}}\ell^2 \left(-\bar{Q}_S\right)^{\frac{1}{2}}\bar{Q}_J. \label{eq:sjj_bound}
\end{equation}
The data is filtered solely by the flux threshold, with the bounding line then overlaid. The close agreement between the data boundary and the theoretical bound demonstrates that the filtered data naturally saturates the bound. 

To visualise the flux ceiling, we fix a maximum value for one or more invariants, $Q\le Q_\mathrm{max}$. Any point satisfying this condition has a theoretical bound no greater than that evaluated at $Q_{\mathrm{max}}$. Consequently, when plotting the flux against the remaining invariants, this bound acts as a ceiling and every flux value in the filtered set lies below it.
This is demonstrated for $\Pi^{I, \ell}_{\mathrm{L}, \Omega S \Omega}$, where the data is restricted to $\bar{Q}_\Omega \le \langle\bar{Q}_\Omega\rangle$ and the corresponding bound overlaid. From Eq.~\eqref{eq:invariant_bound} and Table~\ref{tab:results}, this bound is given by 
\begin{equation}
    |\Pi^{I, \ell}_{\mathrm{L},\Omega S\Omega}| \le \ell^2 \frac{2}{\sqrt{3}} \left(-\bar{Q}_S\right)^{1/2}\left(\bar{Q}_\Omega\right) \label{eq:oso_bound}
\end{equation}
In both cases, the theoretical bounds are seen to be obeyed and tight.

\begin{figure}[ht!]
     \centering
     \begin{subfigure}[b]{0.48\textwidth}
        \centering
        \includegraphics[width=1\linewidth]{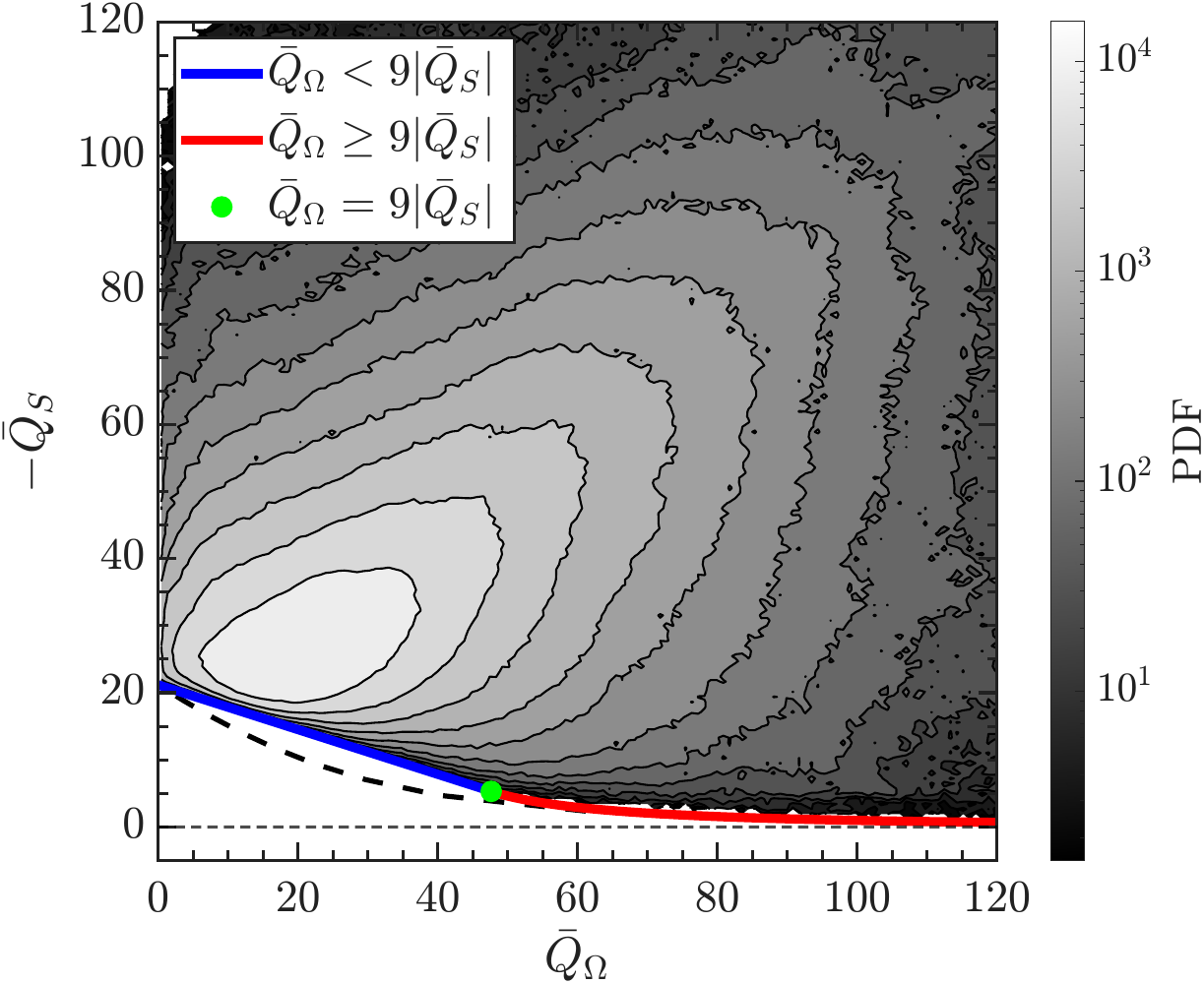}
        \caption{}
        \label{fig:inertial_bound}
     \end{subfigure}
     \begin{subfigure}[b]{0.39\textwidth}
        \centering
        \includegraphics[width=1\linewidth]{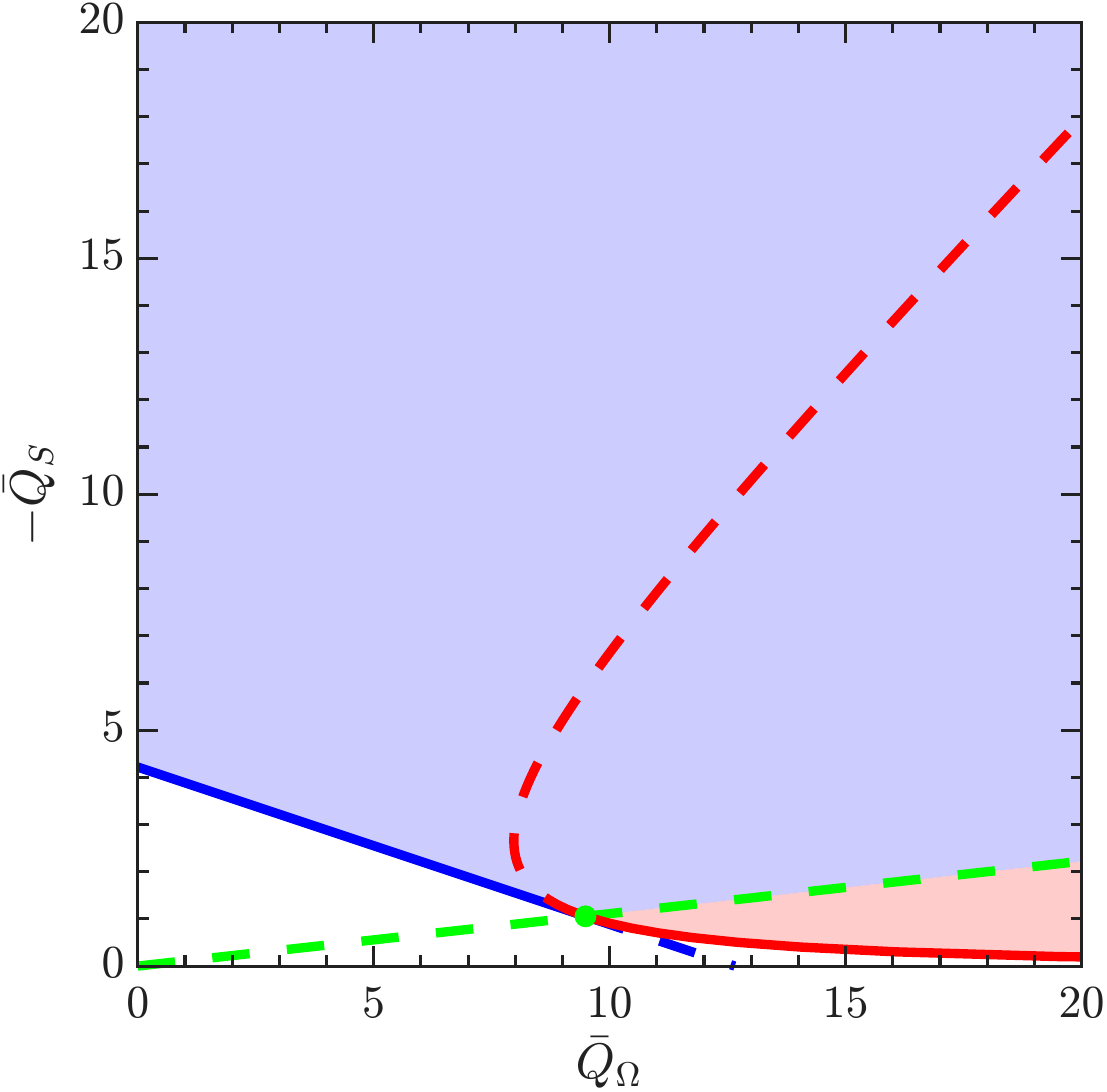}
        \caption{}
        \label{fig:inertial_bound_schematic}
     \end{subfigure}

     \caption{(a) Joint PDF of second invariants $(\bar{Q}_\Omega, -\bar{Q}_S)$ with data filtered by the local Inertial energy flux $|\Pi^{I, \ell}_{\mathrm{L}}|\ge 0.5$, with the corresponding bound Eq.~\eqref{eq:inertial_bound} plotted as blue dash for $\bar{Q}_\Omega < 9|\bar{Q}_S|$, red dash for $\bar{Q}_\Omega \ge 9|\bar{Q}_S|$, and the point of intersection (bifurcation point) as the green marker. The black dashed line is the bound given by the triangle inequality. Simulation data filtered at scale $\ell\approx47\eta$. (b) Schematic of energy flux bound in Eq.~\eqref{eq:inertial_bound} for filter scale $\ell= 1$. The bounded region for $\bar{Q}_\Omega < 9|\bar{Q}_S|$ is shaded light blue, and the bounded region for $\bar{Q}_\Omega \ge 9|\bar{Q}_S|$ is shaded light red. }
     \label{fig:placehodler}
\end{figure}

While repeated application of the triangle inequality yields valid upper bounds for the combined fluxes, these are generally not realisable because the constituent terms cannot simultaneously saturate their individual limits. By formulating the exact optimisation problem, we find that the least upper bound for the Inertial term features a bifurcation at $\bar{Q}_\Omega = 9|\bar{Q}_S|$, 
\begin{equation} 
    |\Pi^{I, \ell}_{\mathrm{L}}| \le \begin{cases}
        \frac{2}{\sqrt{3}}\ell^2  \left(\bar{Q}_\Omega + \bar{Q}_S \right) \left(-\bar{Q}_S \right)^{1/2} & \text{if } \bar{Q}_\Omega \ge 9|\bar{Q}_S|, \\
        \frac{2}{9}\ell^2(\bar{Q}_\Omega - 3\bar{Q}_S)^{3/2} & \text{if } \bar{Q}_\Omega < 9|\bar{Q}_S|. 
    \end{cases}\label{eq:inertial_bound}
\end{equation}
The crossover at $\bar{Q}_\Omega = 9|\bar{Q}_S|$ marks the transition between a triaxial strain-dominated configuration and an axisymmetric vortex-tube configuration as the tightest constraint on the energy flux (Derivation in Appendix~\ref{app:bifurcation_derivation}).
Whilst the bound in Eq.~\eqref{eq:sss_bound} is essentially a re-statement of the fact that $\bar{\bm{S}}$ has three real eigenvalues, the bound in Eq.~\eqref{eq:inertial_bound} is non-trivial in the sense that its form is not the consequence of an eigenvalue restriction on the velocity gradient tensor $\bar{\bm{A}}$ alone. This bifurcation behaviour is shown in Fig.~\ref{fig:inertial_bound}, with a schematic of the bound in Fig.~\ref{fig:inertial_bound_schematic} included to aid interpretation. The filtered data closely follows the appropriate branch in each region. Also shown is the weaker bound obtained via the triangle inequality, $|\Pi^{I, \ell}_{\mathrm{L}}| \le |\Pi^{I, \ell}_{\mathrm{L}, SSS}| + |\Pi^{I, \ell}_{\mathrm{L}\Omega S \Omega}|$ (dashed black line), which is visibly less tight than the optimised bound.   

We close this section by noting two physical consequences of the bounds. First, when $\bar{\bm{X}},\bar{\bm{Y}},\bar{\bm{Z}} \in [\bar{\bm{S}}, \bar{\bm{J}}]$, the large-scale dissipation rates directly limit the available energy flux via Eq.~\eqref{eq:invariant_dissipation}. For example, Eq.~\eqref{eq:sss_bound} becomes
\begin{equation}
    |\Pi^{I, \ell}_{\mathrm{L}, SSS}| \le \frac{1}{4\sqrt{3}}\ell^2\left(\frac{D_u}{\nu}\right)^{3/2}.
\end{equation}
In the inertial range we know that $\mathcal{D}_u \ll \Pi$  \citep{aluie2017}, so the energy flux greatly exceeds the dissipation rate, yet remains bounded above by it via the factor $(\mathcal{D}_u/\nu)^{3/2}$ in the example above. This is not contradictory since, for small $\nu$, this factor can remain large even when $\mathcal{D}_u$ itself is small.
Second, when $\bar{\bm{X}},\bar{\bm{Y}},\bar{\bm{Z}} \in [\bar{\bm{S}}, \bar{\bm{\Omega}}]$, Eq.~\eqref{eq:invariant_bound} must be dimensionally consistent with Kolmogorov scaling in the infinite Reynolds number limit. Since $Q_X \sim |\nabla \bar{\bm{u}}|^2$, the bound scales as
\begin{align}
    |\Pi| \sim  \ell^2 \left(|\nabla \bar{\bm{u}}|\right)^{3} \sim  \frac{\delta u_\ell^3}{\ell}\sim \varepsilon, \notag
\end{align}
recovering the Kolmogorov scaling in configuration space, as required.

\subsection{Energy Flux Proxy }\label{sec:proxy}

\subsubsection{Hydrodynamic Case} \label{sec:hydrodynamic_case}
\noindent The hydrodynamic terms in the Inertial flux are composed of single field contributions and may therefore be written explicitly as functions of the invariants. In particular, the flux of kinetic energy due to strain self-amplification and vorticity stretching are fully captured by the third invariants of the velocity field through
\begin{align}
      \Pi^{I, \ell}_{\mathrm{L}, SSS} &=     3\ell^2 \bar{R}_S, \label{eq:sss_invariant_form}\\
    \Pi^{I, \ell}_{\mathrm{L}, \Omega S \Omega} &= \ell^2(\bar{R}_S - \bar{R}_A). \label{eq:oso_invariant_form} 
\end{align}
These follow immediately from the definitions of the tensor invariants (see Sec.~\ref{sec:invariants}). In hydrodynamics, therefore, the energy flux is simply a function of the third invariants of $\bar{A}_{ij}$ and $\bar{S}_{ij}$. In other words, the energy flux is uniquely determined from the structure of the underlying flow as captured by the tensor invariants.  

We remark that substituting Eq.~\eqref{eq:sss_invariant_form} into the bound Eq.~\eqref{eq:sss_bound} gives
\begin{equation}
|\bar{R}_S| \leq \frac{2}{3\sqrt{3}}\left(-\bar{Q}_S\right)^{3/2}, \label{eq:strain_tensor_bound}
\end{equation}
recovering the condition $\Delta = \frac{27}{4} \bar{R}_S^2 + \bar{Q}_S^3 \leq 0$, where $\Delta$ is the discriminant of the characteristic polynomial of $\bar{\bm{S}}$. This must be satisfied for any real symmetric matrix, so in this case the energy flux bound is identical to the fundamental constraint on the strain-rate tensor.

\subsubsection{General Case} \label{sec:general_flux_invariants}
\noindent The general energy flux term $\Pi^{X, \ell}_{\mathrm{L}, SYZ}$ cannot be expressed uniquely in terms of the invariants, since it may contain mixed-field derivatives. However, the flow strain-rate invariants are related to the general energy flux under the conditions set out below. 
In Section \ref{sec:strain_rate_third_invariant}, it was shown that $\bar{R}_S$ takes the sign of, and is proportional to, the intermediate eigenvalue $\lambda_2$.  In Section \ref{sec:flux_tensor_angle_form}, it was shown that the behaviour of a general local flux is governed by the empirical distribution of the directional cosines, $\cos(\psi_{ij})$.
In particular, in the case of a strong preferential alignment with the intermediate eigenvector, $\cos(\psi_{2j}) \approx 1$, the flux is well-approximated by
\begin{equation*}
\Pi^{X, \ell}_{\mathrm{L}, SYZ} \approx C\lambda_2 \mu_j.
\end{equation*}
Consider the particular case for the flux $\Pi^{M, \ell}_{\mathrm{L}, SJJ}$. From Eq.~\eqref{eq:sjj_angle_form}, we have $\mu_j = \frac{1}{4}|\bar{\bm{j}}|^2$ and $C = -\ell^2$, so 
\begin{equation*}
\Pi^{M, \ell}_{\mathrm{L}, SJJ} \approx -\frac{1}{4} \ell^2|\bar{\bm{j}}|^2\lambda_2.
\end{equation*}
Thus the intermediate eigenvalue $\lambda_2$ will determine the sign of the flux and be proportional to its magnitude. In this case, both $\bar{R}_S$ and $\Pi^{M, \ell}_{\mathrm{L}, SJJ}$ have the same dependence on $\lambda_2$, and so $\bar{R}_S$ will be a proxy for the energy flux. The stronger the preferential alignment with the intermediate eigenvector, i.e. the closer $\cos^2(\psi_{i2})$ is to unity, the better $\bar{R}_S$ becomes as a predictor of the local energy flux. Since the magnitude of the flux is proportional to the intermediate eigenvalue, we can use observations (i)-(iii) from Vi\`etes formula for the eigenvalues (cf.~\ref{sec:vietes_formula}) to inform the expected energy flux in $(\bar{R}_S, \bar{Q}_S)$ space.
That is, we expect the magnitude of the energy flux to scale with $(-\bar{Q}_s)^{1/2}$, and the flux to take its extrema on the discriminant line $\bar{R}_S = \pm \frac{2}{3\sqrt{3}}(-\bar{Q}_S)^{3/2}$ for a fixed $\bar{Q}_S$.
We further expect the magnitude to be smallest near the $\bar{R}_S = 0$ line. \emph{These relationships constitute what we call the invariant-flux proxy.} This is our second main result. 

\begin{figure}[ht!]
     \centering
     \begin{subfigure}[b]{0.48\textwidth}
        \centering
        \includegraphics[width=\textwidth]{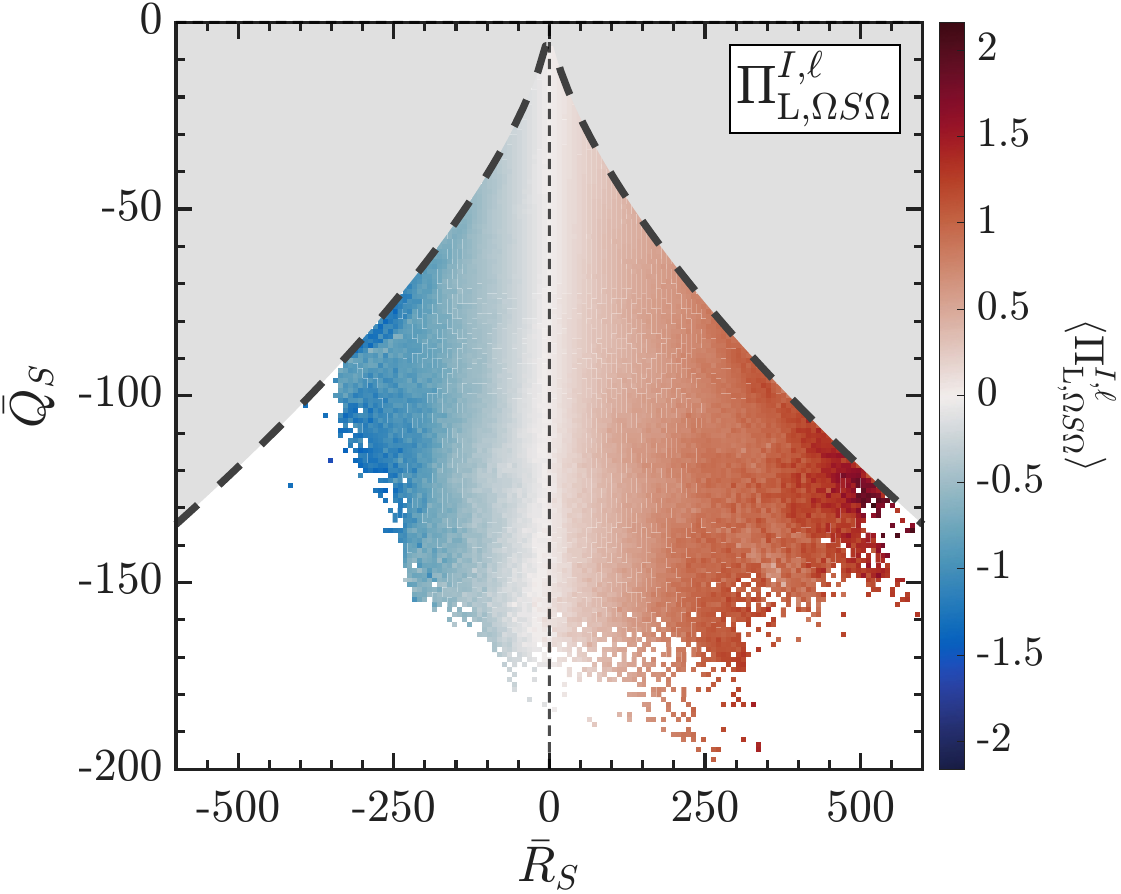}
        \caption{}
        \label{fig:oso_avg}
     \end{subfigure}
     \begin{subfigure}[b]{0.48\textwidth}
        \centering
        \includegraphics[width=\textwidth]{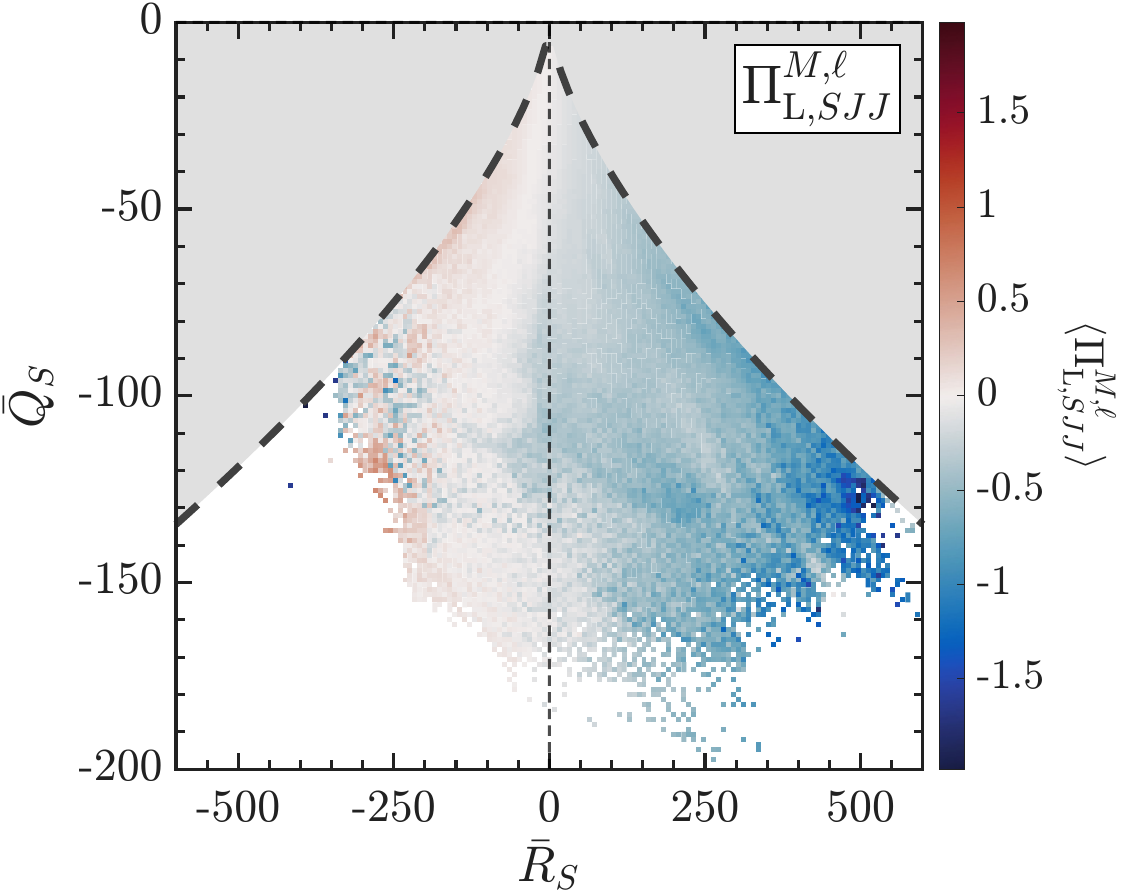}
        \caption{}
        \label{fig:sjj_avg}
     \end{subfigure}
     \caption{Ordering of averaged energy fluxes in the flow strain-rate invariant space $(\bar{R}_S, \bar{Q}_S)$ for (a) energy flux due to vortex stretching $\Pi^{I, \ell}_{\mathrm{L}, \Omega S \Omega}$ and (b) energy flux due to current filament stretching $\Pi^{M, \ell}_{\mathrm{L}, SJJ}$ filtered at scale $\ell \approx 47 \eta$. The colour bars are shown on their natural scales to emphasise the inherent difference in magnitude between the two fluxes. Red and blue colour scale indicate positive and negative energy fluxes respectively. The thick dashed black line is the discriminant line $\bar{R}_S = \pm \frac{2}{3\sqrt{3}}(-\bar{Q}_S)^{3/2}$ (cf. Eq.~\eqref{eq:strain_tensor_bound}). Simulation details in Appendix~\ref{app:simulations}. }

     \label{fig:oso_sjj_avg}

\end{figure}

\begin{figure}[ht!]
     \centering
    \begin{subfigure}[b]{0.48\textwidth}
     \hspace{-1.24cm}
        \centering
        \includegraphics[width=0.78\textwidth]{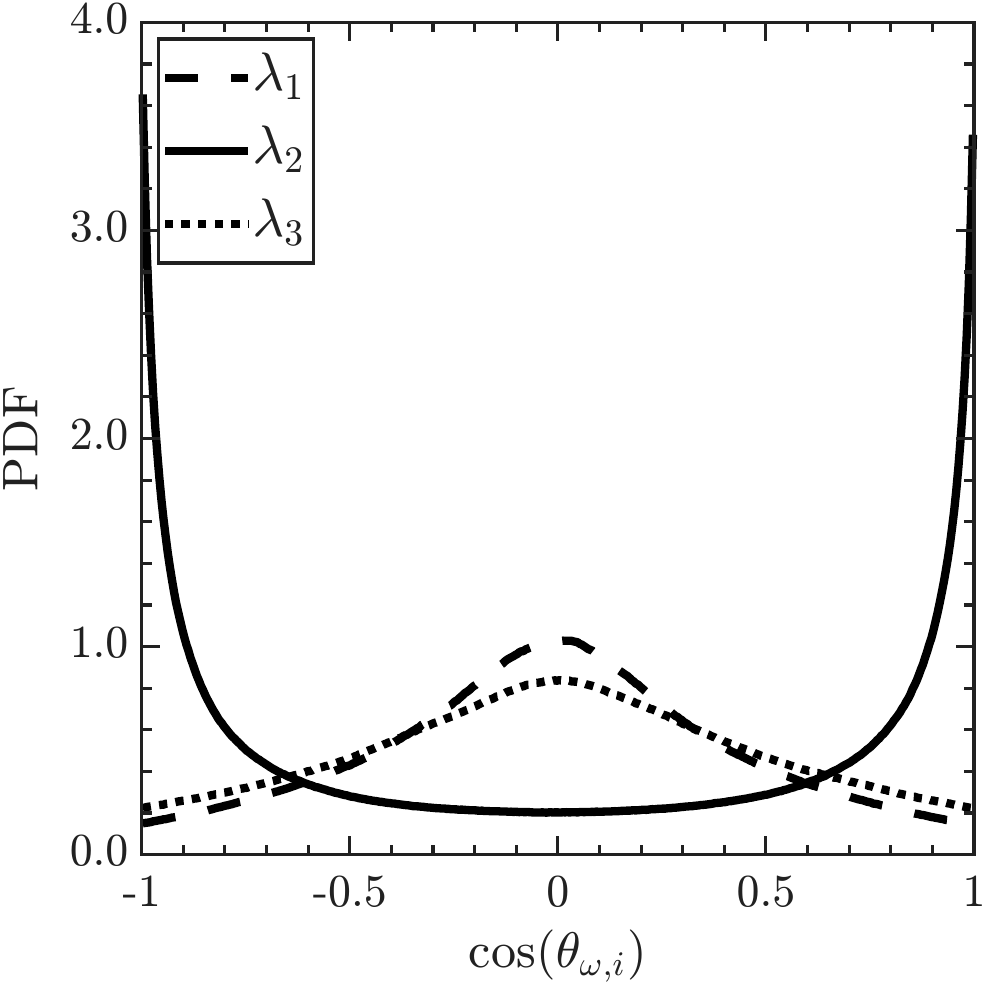}
        \caption{}
        \label{fig:oso_cosine}
     \end{subfigure}
     \begin{subfigure}[b]{0.48\textwidth}
     \hspace*{-1.22cm}
        \centering
        \includegraphics[width=0.78\textwidth]{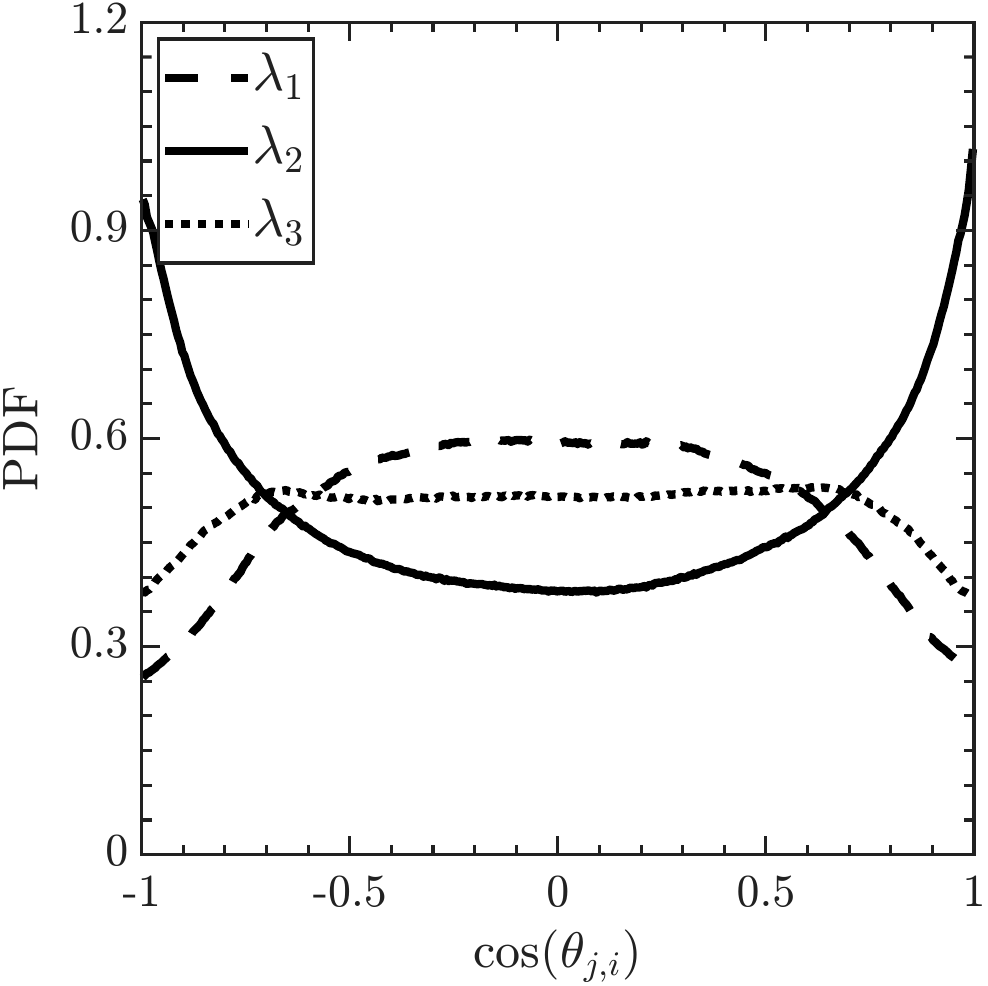}
        \caption{}
        \label{fig:sjj_cosine}
        
     \end{subfigure}

     \caption{PDFs of the alignment cosines of the angle $\theta$ between the eigenvectors (with eigenvalues $\lambda_i$) of the strain-rate tensor and (a) vorticity (b) current density, both showing preferential alignment with the intermediate eigenvector (eigenvalue $\lambda_2$). Simulation details in Appendix~\ref{app:simulations}.}
     
\end{figure}

To illustrate this, we examine the statistics of $\Pi^{I, \ell}_{\mathrm{L}, \Omega S \Omega}$ and $\Pi^{M, \ell}_{\mathrm{L}, SJJ}$, both of which admit the simplified tensor-angle form (see Section \ref{sec:flux_tensor_angle_form}). In this form, the flux depends only on the alignment of the vorticity and current density vectors with the strain-rate eigenvectors, reducing the statistical description from nine directional cosines to three. 
Fig.~\ref{fig:oso_sjj_avg} shows the mean energy flux in strain-rate invariant space $(\bar{R}_S, \bar{Q}_S)$ for the vorticity stretching mechanism (Fig.~\ref{fig:oso_avg}) and the current filament stretching mechanism (Fig.~\ref{fig:sjj_avg}), at filter scale  $\ell\approx47\eta$ corresponding to the middle of the inertial range, where $\eta$ is the Kolmogorov dissipation scale (see Appendix~\ref{app:simulations}). The sign of  $\langle \Pi^{I, \ell}_{\mathrm{L}, \Omega S \Omega} \rangle$ in Fig.~\ref{fig:oso_avg} follows that of $\bar{R}_S$, the flux is weakest near $\bar{R}_S=0$, the strongest values occur near the discriminant line, and the magnitude grows with increasing $-\bar{Q}_S$. This is precisely the behaviour predicted by the framework above, under the condition of strong preferential alignment of the vorticity with the intermediate strain-rate eigenvector. That this condition is well-satisfied is confirmed by Fig.~\ref{fig:oso_cosine}, which shows the PDFs of the alignment cosines between the vorticity and the strain-rate eigenvectors. The vorticity aligns strongly with the intermediate eigenvector associated with $\lambda_2$, while alignment with the extensive and compressive eigenvectors is considerably weaker. This behaviour is well-documented in hydrodynamic turbulence \citep{tsinober2001, davidson2004} and has also been observed \textit{in situ} in MHD turbulence \citep{consolini2015}.

The current filament stretching term in Fig.~\ref{fig:sjj_avg} displays the opposite sign convention, with $\langle \Pi^{M, \ell}_{\mathrm{L}, SJJ} \rangle$ following the opposite sign of $\bar{R}_S$ owing to the negative prefactor in Eq.~\eqref{eq:sjj_angle_form}. The ordering of the flux with the invariants remains visible but is notably weaker than for the vorticity stretching term, suggesting poorer geometric alignment. This is confirmed by Fig.~\ref{fig:sjj_cosine}, which shows that whilst the current density still tends to align with the intermediate eigenvector, the preference is weaker and the contributions from the extensive and compressive eigenvectors are more significant. Consequently, localised violations of the invariant ordering are expected and indeed visible in Fig.~\ref{fig:sjj_avg}.

The result above depends on the convenient property that in the approximation to the general flux, $\Pi^{X, \ell}_{\mathrm{L}, SYZ} \approx C\lambda_2 \mu_j$, $\mu_j$ is a constant, independent of the local strain-rate topology. In general, the local flux depends on $\mu_j$. If this fluctuates unpredictably for a given $(\bar{R}_S, \bar{Q}_S)$, the invariants cease to describe the energy flux accurately. There are two ways to overcome this restriction and thereby generalise the result: (i) if $\mu_j$ is statistically independent of the local strain-rate topology or (ii) if $\mu_j$ is a known function of $(\bar{R}_S, \bar{Q}_S)$.
Condition (ii) is simply the statement that the local energy flux, $ \Pi \approx  C\lambda_2 \mu_j$, is a known function of $(\bar{R}_S, \bar{Q}_S)$, and hence the invariants are proxies for the flux under preferential alignment. 
Condition (i) allows us to make strong statements about the conditional average of the general energy flux $\Pi^{X,\ell}_{Y, Z}$ in the $(\bar{R}_S, \bar{Q}_S)$ plane,
\begin{equation}
    \langle \Pi^{X,\ell}_{Y, Z} | (\bar{R}_S, \bar{Q}_S)\rangle \approx  C\lambda_2\langle \mu_j | (\bar{R}_S, \bar{Q}_S)\rangle. 
\end{equation}
The eigenvalue $\lambda_2$ is outside the expectation since it is determined exactly by $(\bar{R}_S, \bar{Q}_S)$. Under condition (i), the conditional average $\langle \mu_j | (\bar{R}_S, \bar{Q}_S)\rangle$ is simply the global average, $\langle \mu_j \rangle$, a constant in $(\bar{R}_S, \bar{Q}_S)$ space. The averaged flux in invariant space is then approximated by
\begin{equation}
    \langle \Pi^{X,\ell}_{Y, Z} | (\bar{R}_S, \bar{Q}_S)\rangle \approx  C\lambda_2\langle \mu_j\rangle, 
\end{equation}
which has the same form as the $\Pi^{M, \ell}_{\mathrm{L}, SJJ}$ term and hence the relationship with the strain-rate invariants is identical to the discussion above. 

\begin{figure}[ht!]
     \centering
        \includegraphics[width=0.5\textwidth]{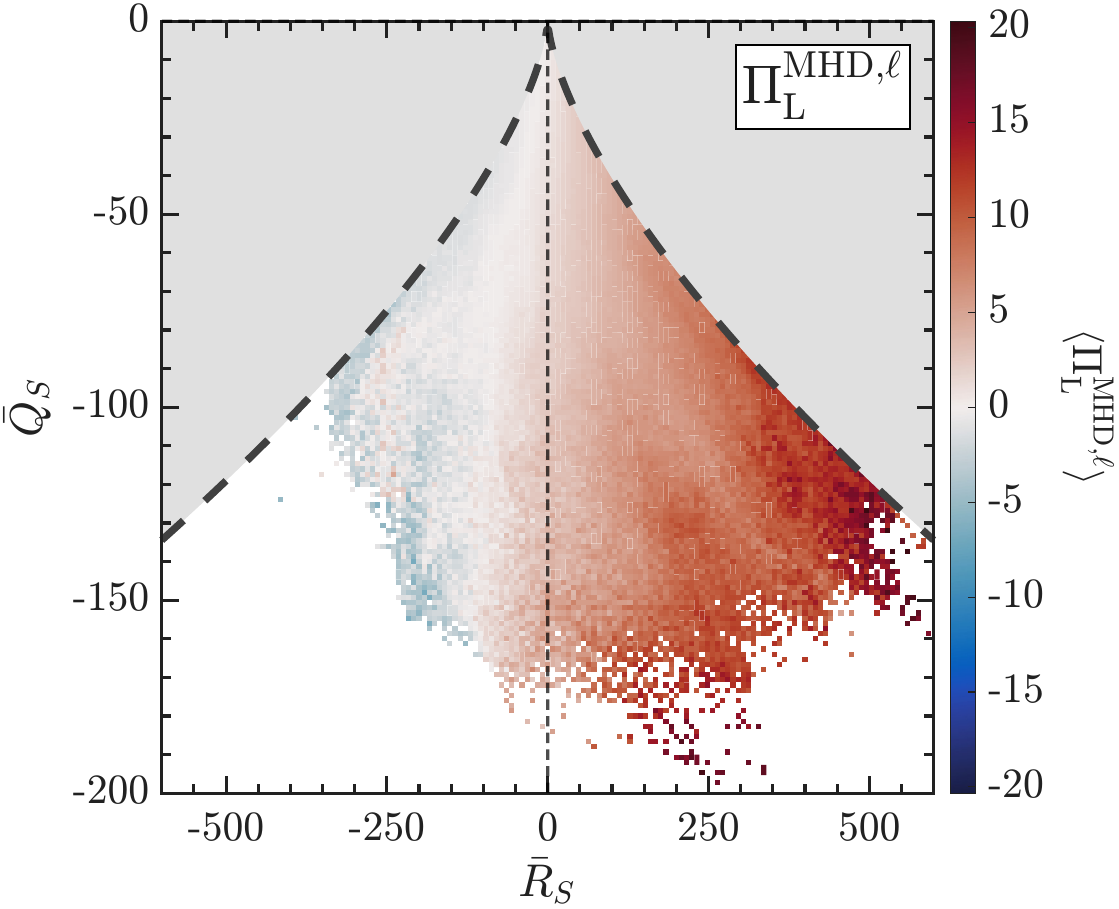}
         \caption{Ordering of averaged local total energy flux in the flow strain-rate invariant space, $\langle\Pi^{\mathrm{MHD}, \ell}_{\mathrm{L}}| (\bar{R_S}, \bar{Q_S})\rangle $. Red and blue correspond to forward and backward energy flux respectively. The thick dashed black line is the discriminant line $\bar{R}_S = \pm \frac{2}{3\sqrt{3}}(-\bar{Q}_S)^{3/2}$ (cf. Eq.~\eqref{eq:strain_tensor_bound}). The minimum count in each bin is 10. Simulation details in Appendix~\ref{app:simulations}.}
         \label{fig:average_flux_strain-rate}
\end{figure}

To illustrate the applicability of this general relationship, Fig.~\ref{fig:average_flux_strain-rate} shows  $\Pi^{\mathrm{MHD}, \ell}_{\mathrm{L}}$ averaged in $(\bar{R}_S, \bar{Q}_S)$ space. Despite $\Pi^{\mathrm{MHD}, \ell}_{\mathrm{L}}$ being a general flux term not restricted to the simplified vector-angle form, the averaged flux displays the same ordering with the strain-rate invariants as before: the sign follows that of $\bar{R}_S$, the magnitude grows with $-\bar{Q}_S$, and the strongest values occur near the discriminant line. This confirms that conditions (i)-(ii) are well-satisfied on average for the constituent fluxes, so that the strain-rate invariants serve as reliable proxies for the general energy flux under conditional averaging.

\begin{figure}[ht!]
    \centering
    \includegraphics[width=1\textwidth]{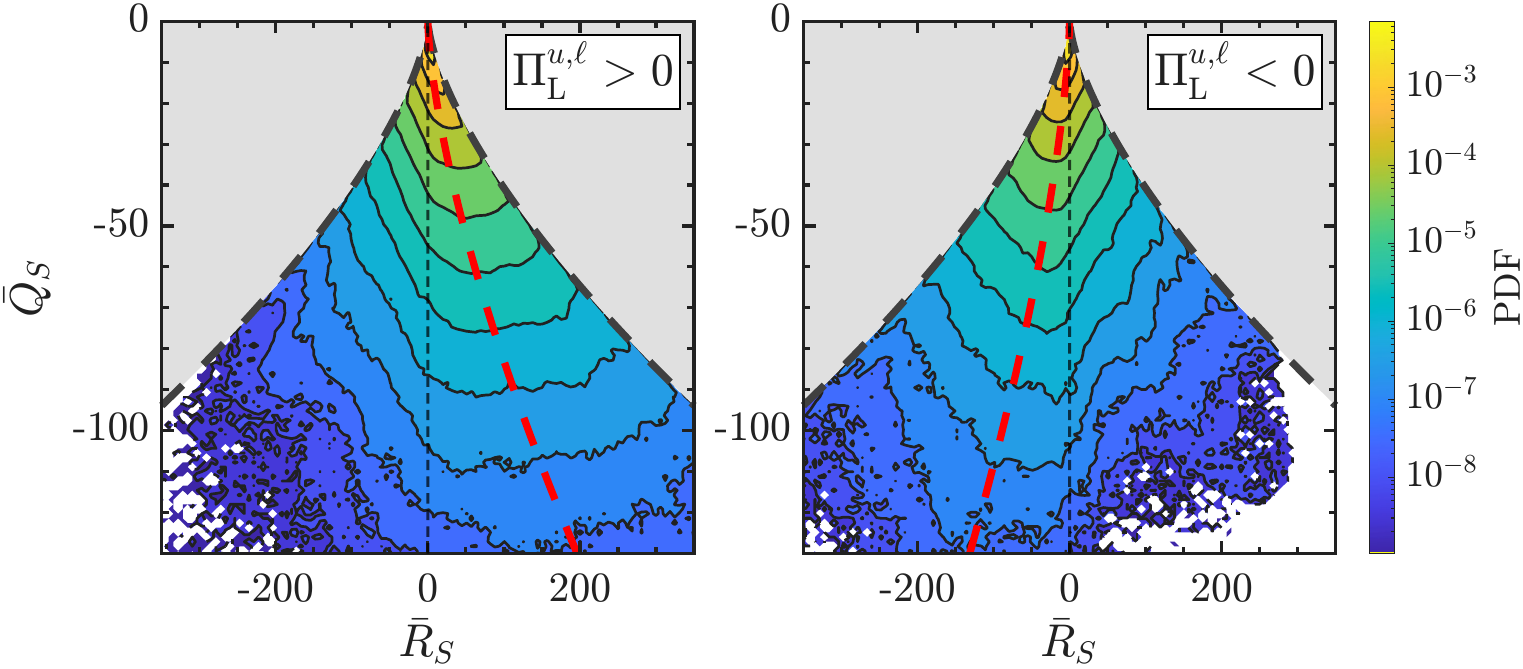}
    \caption{Joint PDF of the flow strain-rate invariants $(\bar{R}_S, \bar{Q}_S)$ conditioned on the sign of local kinetic energy flux $\Pi^{u, \ell}_{\mathrm{L}}$. Red dashed lines correspond to eigenvalue ratio lines given by Eq.~\eqref{eq:eig_ratio}. The thick dashed black line is the discriminant line $\bar{R}_S = \pm \frac{2}{3\sqrt{3}}(-\bar{Q}_S)^{3/2}$ (cf. Eq.~\eqref{eq:strain_tensor_bound}). Simulation details in Appendix~\ref{app:simulations}.}
    \label{fig:strain_rate_kinflux_15}
\end{figure}

It is also instructive to examine the PDFs of the strain-rate invariants conditioned on the sign of the local energy flux, which reveals how forward and inverse energy transfer is linked to the flow structures described by the $(\bar{R}_S, \bar{Q}_S)$ plane. Fig.~\ref{fig:strain_rate_kinflux_15} shows these conditioned PDFs for the local kinetic energy flux $\Pi^{I, \ell}_{\mathrm{L}}$. There is a strong correlation between the sign of the flux and the sign of $\bar{R}_S$, confirming that the third strain-rate invariant is a reliable predictor of the direction of local energy transfer, consistent with the framework developed above. The red dashed lines show the averaged eigenvalue lines, given by
\begin{equation}
    \bar{R}_S = (-\bar{Q}_S)^{3/2}a(1+a)(1+a+a^2)^{-3/2}, \quad \mathrm{where} \quad a= \frac{\lambda_2}{\lambda_1}. \label{eq:eig_ratio}
\end{equation}
where each line corresponds to a distinct flow topology (see \citet{dallasStructuresDynamicsSmall2013} for details), with $a$ evaluated using the conditional averages $a = \langle \lambda_2\rangle/\langle \lambda_1\rangle$. These lines allow us to assess whether the structures driving the forward and inverse cascade are qualitatively different. For the forward cascade of kinetic energy in Fig.~\ref{fig:strain_rate_kinflux_15}, the mean eigenvalue ratios are $\langle\lambda_1 \rangle : \langle\lambda_2 \rangle : \langle\lambda_3 \rangle \approx 1: 0.15 : -1.15$, consistent with the simulation results of \citet{dallasStructuresDynamicsSmall2013} and reasonably close to the universal $3:1:-4$ ratio reported for hydrodynamic turbulence \citep{tsinober2001} for the bulk of the distribution. For the inverse cascade in Fig.~\ref{fig:strain_rate_kinflux_15}, the ratios are $\langle\lambda_1 \rangle : \langle\lambda_2 \rangle : \langle\lambda_3 \rangle =  1: -0.09 : -0.91$. The sign change in $\langle \lambda_2\rangle$ is the key distinction, with the forward cascade associated with biaxial stretching structures, whilst the inverse cascade is associated with biaxial compression. In both cases the ratios lie closer to the two-dimensional flow limit ($\bar{R}_S=0$) than is typical of hydrodynamic turbulence.

\section{Discussion}

This work has established two complementary relationships between the tensor invariants and the scale-filtered energy flux in MHD turbulence. First, the energy flux is bounded above by a function of the local invariants, with the least upper bound obtained by solving the exact optimisation problem. Second, under conditions of preferential alignment and statistical independence of $\mu_j$ from the local strain-rate topology, the strain-rate invariants serve as proxies for the local energy flux, with the sign of the third invariant $\bar{R}_S$ being a reliable predictor of the direction of energy transfer.
Both results were confirmed against simulation data, with the theoretical bounds shown to be tight and the conditional averaged flux in $(\bar{R}_S, \bar{Q}_S)$ space displaying the predicted ordering.

The most immediate application of these results is to multispacecraft observations, where the tensor invariants can be estimated from the velocity and magnetic field gradients computed across spacecraft separations at scale $\ell$ \citep{paschmannAnalysisMethodsMultiSpacecraft, dentonPolynomialReconstructionReconnection2020}. The invariant statistics interpreted in this context provide an accessible route to characterising energy transfer. In particular, the proxy relationship established here means that the conditional statistics of the invariants can be used in place of direct flux estimates, and a comparison with existing cascade rate estimators, such as the local energy transfer (LET) method \citep{sorriso-valvoLocalEnergyTransfer2018, sorriso-valvo2019}, would provide a consistency check.

A practical consideration for such an application concerns the filter kernel. The effective filter implicit in interpolating gradients across spacecraft positions is not Gaussian as used here. However, the local flux term is the leading-order term in the power-law expansion of any filter kernel with finite moments in the limit of small filter scale \citep{pope2015}. The results derived here are therefore applicable to multispacecraft measurements.

Several limitations of the present study should be noted. The simulations were conducted for illustrative purposes in a relatively idealised setting. Tailoring the numerical setup to specific physical environments would yield more quantitative predictions. For example, modelling the solar wind would require the inclusion of a background magnetic field, and the effects of compressibility, extended MHD, and anisotropy would remain to be addressed. The present analysis was also performed at a fixed filtering scale in the inertial range, and a systematic study of how the conditional flux statistics and the tightness of the bounds vary across the inertial range and into the dissipation range would be a natural extension of this work. 


\begin{acknowledgments}
This work was partially supported by RCUK grant CG ST/X000915/1. We acknowledge AFOSR grant FA8655-22-1-7056. The authors thank Prof. P. Mininni and the developers of the GHOST code for providing the code used to run the simulations presented in this work.
\end{acknowledgments}

\appendix

\section{Simulation Details} \label{app:simulations}
\begin{figure}[ht!]
    \centering
    \includegraphics[width=0.85\linewidth]{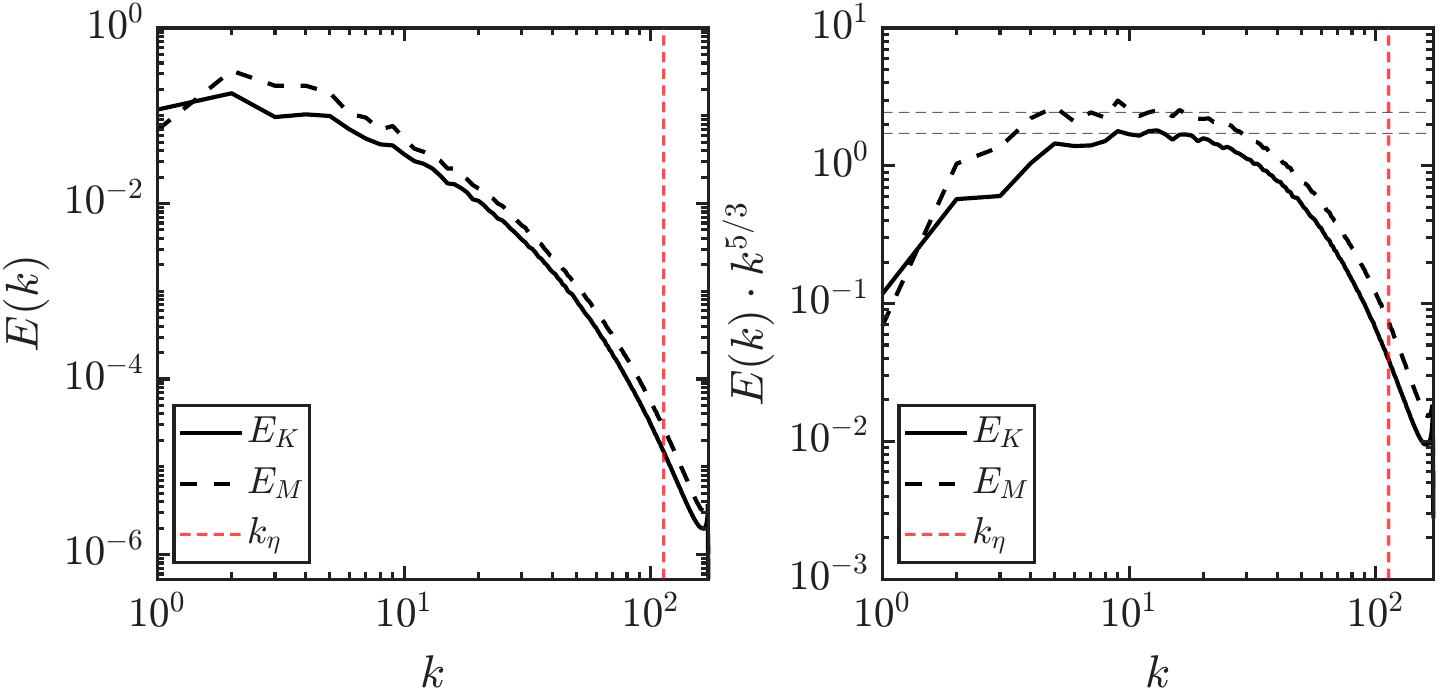}
    \caption{Simulation energy spectra for kinetic energy (solid) and magnetic energy (dashed), plotted (left) log-log and (right) compensated by the Kolmogorov scaling. On both plots the red vertical line is the Kolmogorov dissipation wavenumber.}
    \label{fig:spectra}
\end{figure}

\noindent We perform 3D freely decaying incompressible MHD simulations using the GHOST pseudospectral code \citep{gomez2005}. The simulations are conducted in a periodic cubic domain of size $[0, 2\pi]^3$, with random-phase initial conditions \citep{pouquetNumericalSimulationHelical1978}. 
A ratio of $k_{\mathrm{max}}/k_\eta>1.5$ is maintained to ensure sufficient resolution of the dissipative scales, corresponding to a viscosity and magnetic diffusivity of $\nu = \eta = 1.6 \times 10^{-3}$. 
Analysis is performed at the time of maximum dissipation when the turbulence is in a fully developed state. The simulation is ran without a background field, $B_0 = 0$. 
Fig.~\ref{fig:spectra} shows the isotropic magnetic and kinetic energy spectra and the same spectra compensated by the Kolmogorov scaling $k^{-5/3}$. The Iroshnikov–Kraichnan scaling $k^{-3/2}$ (not shown) exhibits a less extended inertial range. The simulation results in the main text are filtered in the middle of the inertial range, corresponding to wavenumber $k\approx15$ from the spectra. The filter scale in units of the Kolmogorov dissipation scale $\eta$ is then
\begin{equation*}
    \frac{\ell}{\eta} \approx \frac{2\pi/k}{1.5/k_{\mathrm{max}}}\approx47.
\end{equation*}

\section{Mathematical Details}

\subsection{Derivation of Vector-Angle Form} \label{app:vector-angle_form}
Both the current filament stretching and vorticity stretching terms share a general form involving the triple contraction of the symmetric strain-rate tensor, $\bar{S}_{ij}$, with two anti-symmetric tensors that form the dual tensors for a vector $\bm{x}$. Using the dual tensor relation, $\bar{X}_{ij} = -\frac{1}{2}\epsilon_{ijk}\bar{x}_{k}$, the contraction of the antisymmetric tensors expands as 
\begin{equation*}
    \bar{X}_{ik}\bar{X}_{jk} = \frac{1}{4}\left(|\bar{\bm{x}}|^2 \delta_{ij} - \bar{x}_i\bar{x}_j\right),
\end{equation*}
after invoking the identity $\epsilon_{ikm}\epsilon_{jkn} = \delta_{ij}\delta_{mn} - \delta_{in}\delta_{mj}$. Contracting this with the resolved strain-rate tensor yields
\begin{equation*}
\bar{S}_{ij}\bar{X}_{ik}\bar{X}_{jk} = \frac{1}{4}|\bar{\bm{x}}|^2\bar{S}_{ii} - \frac{1}{4}\bar{S}_{ij}\bar{x}_i\bar{x}_j.
\end{equation*}
The first term vanishes due to incompressibility, leaving the quadratic form
\begin{equation*}
    \bar{S}_{ij}\bar{X}_{ik}\bar{X}_{jk} = -\frac{1}{4}\bar{S}_{ij}\bar{x}_i\bar{x}_j = -\frac{1}{4} \bar{\bm{x}}^T \bar{\bm{S}} \bar{\bm{x}}.
\end{equation*}
Because the strain-rate tensor $\bar{\bm{S}}$ is real and symmetric, it can be diagonalised as $\bar{\bm{S}} = \bm{Q}\bm{\Lambda}\bm{Q}^T$, where $\bm{\Lambda} = \mathrm{diag}(\lambda_1, \lambda_2, \lambda_3)$ is the diagonal matrix of real eigenvalues and $\bm{Q}$ is an orthogonal matrix whose columns are the corresponding orthonormal eigenvectors. Substituting this decomposition into the quadratic form gives
\begin{equation*}
    \bar{\bm{x}}^T \bar{\bm{S}} \bar{\bm{x}} = \bar{\bm{x}}^T (\bm{Q}\bm{\Lambda}\bm{Q}^T) \bar{\bm{x}} = (\bm{Q}^T\bar{\bm{x}})^T \bm{\Lambda} (\bm{Q}^T\bar{\bm{x}}).
\end{equation*}
Let the transformed vector be denoted  $\bar{\bm{x}}' = \bm{Q}^T\bar{\bm{x}}$. Because $\bm{Q}$ is an orthogonal matrix, this operation represents a pure spatial rotation into the eigenbasis of the strain-rate tensor that preserves the vector magnitudes, $|\bar{\bm{x}}'| = |\bar{\bm{x}}|$. The $i$-th component of the rotated vector in this eigenbasis is then the projection of $\bar{\bm{x}}$ onto the $i$-th eigenvector,
\begin{equation*}
    \bar{x}'_i = |\bar{\bm{x}}|\cos(\theta_{x,i}),
\end{equation*}
where $\theta_{x,i}$ is the angle between the vector $\bar{\bm{x}}$ and the $i$-th eigenvector of $\bar{\bm{S}}$. The quadratic form then reduces to a scalar sum over the eigenvalues, 
\begin{equation*}
    \bar{\bm{x}}^T \bar{\bm{S}} \bar{\bm{x}} = \bar{\bm{x}}^T \bm{\Lambda} \bar{\bm{x}}' = |\bar{\bm{x}}|^2 \sum_{i=1}^3 \lambda_i \cos^2(\theta_{x,i}).
\end{equation*}
Reintroducing the appropriate constants we obtain the exact vector-angle formulations for the vorticity stretching energy flux,
\begin{equation}
    \Pi^{I, \ell}_{\mathrm{L}, \Omega S \Omega} = \frac{1}{4} \ell^2|\bar{\bm{\omega}}|^2\sum_{i=1}^3\lambda_i\cos^2(\theta_{\omega,i}), 
    \label{eq:oso_tensor_angle}
\end{equation}
and the current filament stretching energy flux,
\begin{equation*} \tag{\ref{eq:sjj_angle_form}}
    \Pi^{M, \ell}_{\mathrm{L}, SJJ} = -\frac{1}{4} \ell^2|\bar{\bm{j}}|^2\sum_{i=1}^3\lambda_i\cos^2(\theta_{j,i}). 
\end{equation*}
This recovers Eq.~\eqref{eq:oso_tensor_angle} and Eq.~\eqref{eq:sjj_angle_form} (main text) as required. We note that in the incompressible case, the current filament stretching term would be 
\begin{equation*}
    \Pi^{M, \ell}_{\mathrm{L}, SJJ} = \frac{1}{4} \ell^2|\bar{\bm{j}}|^2 \left(\mathrm{tr}(\bar{\bm{S}}) -\sum_{i=1}^3\lambda_i\cos^2(\theta_{j,i})\right) = \frac{1}{4} \ell^2|\bar{\bm{j}}|^2 \sum_{i=1}^3\lambda_i\left(1-\cos^2(\theta_{j,i})\right),
\end{equation*}
with a similar form for the vorticity stretching term. 

\subsection{Equivalence of General Tensor-Angle form and Vector-Angle Form}\label{app:tensor_angle_vector_angle}

We show that the vector-angle forms of the fluxes $\Pi^{I, \ell}_{\mathrm{L}, \Omega S \Omega}$ and $\Pi^{M, \ell}_{\mathrm{L}, SJJ}$ in equations  Eq.~\eqref{eq:oso_tensor_angle} and Eq.~\eqref{eq:sjj_angle_form} are equivalent to the  general tensor-angle form in Eq.~\eqref{eq:flux_angle_general_form} (main text). Both forms are well-known \citep{ballouz2018, johnson2021, capocciEnergyFluxDecomposition2025}, but their explicit connection has not, to the best of our knowledge, been shown before. We include it here for completeness. This is shown for the vorticity stretching term but the current filament stretching term is mathematically identical. Start with the tensor contraction term in the flux definition, $\bar{S_{ij}} \bar{Y_{ik}} \bar{Z_{jk}}$, and substitute $\bar{\bm{Y}} = \bar{\bm{Z}} = \bar{\bm{\Omega}}$. Let $M_{ij} = \bar{\Omega}_{ik} \bar{\Omega}_{jk}$. We can rewrite this product as a standard matrix multiplication, 
\begin{equation*}
M_{ij} = \Omega_{ik} (-\Omega_{kj}) = -(\bm{\Omega}^2)_{ij}.
\end{equation*}
Since the square of an antisymmetric matrix is symmetric, $\bm{M}$ is symmetric, so the ``symmetric part" required by the definition of Eq.~\eqref{eq:flux_angle_general_form} is the matrix itself. We need the eigenvalues of $\bm{M} = -\bm{\Omega}^2$. 
Using the definition $\Omega_{ij} = -\frac{1}{2}\epsilon_{ijk}\omega_k$, the action of the operator $-\bm{\Omega}^2$ on a vector $\bm{x}$ is
\begin{equation*}
-\bm{\Omega}^2 \bm{x} = \frac{1}{4} \left[ |\bm{\omega}|^2 \bm{x} - (\bm{\omega} \cdot \bm{x})\bm{\omega} \right]
\end{equation*}
By inspection, one eigenvector is parallel to the vorticity, $\bm{u}_1 \parallel \bm{\omega}$, with corresponding eigenvalue $\mu_1 = 0$, since
\begin{equation*}
-\bm{\Omega}^2 \bm{\omega} = \frac{1}{4} [ |\bm{\omega}|^2 \bm{\omega} - |\bm{\omega}|^2 \bm{\omega} ] = 0.
\end{equation*}
The other eigenvectors are then perpendicular to vorticity, $\bm{u}_{2,3} \perp \bm{\omega}$, with eigenvalues $\mu_2 = \mu_3 = \frac{1}{4}|\bm{\omega}|^2$, since
\begin{equation*}
-\bm{\Omega}^2 \bm{x}_{\perp} = \frac{1}{4} |\bm{\omega}|^2 \bm{x}_{\perp}.
\end{equation*}
The tensor-angle form in Eq.~\eqref{eq:flux_angle_general_form} expanded out is 
\begin{equation*}
\Pi^{I, \ell}_{\mathrm{L}, \Omega S \Omega} = \ell^2\sum_{i=1}^3 \lambda_i \left[ \mu_1 \cos^2(\psi_{i1}) + \mu_2 \cos^2(\psi_{i2}) + \mu_3 \cos^2(\psi_{i3}) \right].
\end{equation*}
Substituting these eigenvalues into this sum gives 
\begin{equation*}
\Pi^{I, \ell}_{\mathrm{L}, \Omega S \Omega}= \ell^2 \sum_{i=1}^3 \lambda_i \left[ 0 + \frac{1}{4}|\bm{\omega}|^2 \left( \cos^2(\psi_{i2}) + \cos^2(\psi_{i3}) \right) \right].
\end{equation*}
The angles $\psi_{ij}$ are measured between the $i$-th strain-rate eigenvector and the $j$-th eigenvector vector of $\bm{M}$. Since the eigenbasis of $\bm{M}$ is orthogonal, the sum of squared cosines between the $i$-th strain-rate eigenvector and the eigenvectors of $\bm{M}$ must be unity,
\begin{equation*}
\cos^2(\psi_{i1}) + \cos^2(\psi_{i2}) + \cos^2(\psi_{i3}) = 1.
\end{equation*}
We can then replace the perpendicular terms with the vorticity eigenvector term,
\begin{equation*}
\cos^2(\psi_{i2}) + \cos^2(\psi_{i3}) = 1 - \cos^2(\psi_{i1}). 
\end{equation*}
Here $\psi_{i1}$ is the angle between the $i$-th strain-rate eigenvector and the vorticity vector. Denote this angle $\theta_{i}$ and expand the summation
\begin{equation*}
\Pi^{I, \ell}_{\mathrm{L}, \Omega S \Omega} =  \frac{1}{4}\ell^2|\bm{\omega}|^2 \left( \sum_{i=1}^3 \lambda_i - \sum_{i=1}^3 \lambda_i \cos^2(\theta_{i}) \right).
\end{equation*}
Due to incompressibility, $\sum \lambda_i = 0$, we are left with
\begin{equation*}
\Pi^{I, \ell}_{\mathrm{L}, \Omega S \Omega} = \frac{1}{4}\ell^2|\bm{\omega}|^2 \sum_{i=1}^3 \lambda_i \cos^2(\theta_{i}).
\end{equation*}
This is exactly the vector-angle form in Eq.~\eqref{eq:oso_tensor_angle}. Equation~\eqref{eq:sjj_angle_form} is recovered similarly.

\subsection{Lagrange Multiplier Optimisation: Bifurcation Example} \label{app:bifurcation_derivation}
We derive the bound for the local inertial flux $\Pi^{I, \ell}_{\mathrm{L}} = -\ell^2\mathrm{tr}\left(\bar{\bm{A}}^T\bar{\bm{A}}\bar{\bm{A}}^T\right)$. From the expansion $\mathrm{tr}(\bar{\bm{A}}^T\bar{\bm{A}}\bar{\bm{A}}^T) = \mathrm{tr}(\bar{\bm{S}}^3) - \mathrm{tr}(\bar{\bm{S}}\bar{\bm{\Omega}}^2)$, working in the eigenbasis of $\bar{\bm{S}}$ this second term can be written 
\begin{align*}
    -\mathrm{tr}(\bar{\bm{S}}\bar{\bm{\Omega}}) &= -\mathrm{tr}\left[\begin{pmatrix}
       \lambda_1 & 0 & 0 \\
        0 &\lambda_2 & 0 \\
        0 & 0 &\lambda_3
    \end{pmatrix}
    \begin{pmatrix}
        0 & \Omega_{12} & \Omega_{13} \\
        -\Omega_{12} & 0 & \Omega_{23} \\
        -\Omega_{13} & -\Omega_{23} & 0
    \end{pmatrix}^2\right]\\
    &= (\lambda_1+\lambda_2)\Omega_{12}^2 + (\lambda_1+\lambda_3)\Omega_{13}^2 + (\lambda_2+\lambda_3)\Omega_{23}^2\\
    &= -\lambda_3\Omega_{12}^2 -\lambda_2 \Omega_{13}^2 -\lambda_1\Omega_{23}^2 
\end{align*}
To maximise this sum subject to a fixed $ \bar{Q}_\Omega=\Omega_{12}^2+\Omega_{13}^2 + \Omega_{23}^2$, all vorticity must be concentrated into the component corresponding to the most positive coefficient. Since $\lambda_1+\lambda_2+\lambda_3=0$, we take without loss of generality $\lambda_3$ negative and $\lambda_3 \le \lambda_2 \le \lambda_1$. Then the sum is maximised when $\Omega_{13}^2 = \Omega_{23}^2 = 0$, so that $\bar{Q}_\Omega = \Omega_{12}^2$. This amounts to concentrating vorticity on the $(1,2)$ off-diagonal component, and the objective reduces to
\begin{equation*}
    F = \sum_i\lambda_i^3 - \bar{Q}_\Omega\lambda_3
\end{equation*}
with $\lambda_3$ being the eigenvalue not paired with the vorticity. The Lagrangian is
\begin{equation*}
    \mathcal{L}_s = \sum_i\lambda_i^3 - \bar{Q}_\Omega\lambda_3 - \alpha\sum_i\lambda_i - \beta\left(\sum_i\lambda_i^2 + 2\bar{Q}_S\right).
\end{equation*}
The stationarity conditions are then
\begin{align}
3\lambda_1^2 - \alpha - 2\beta\lambda_1 &= 0, \label{a} \\
3\lambda_2^2 - \alpha - 2\beta\lambda_2 &= 0, \label{b} \\
3\lambda_3^2 - \bar{Q}_\Omega - \alpha - 2\beta\lambda_3 &= 0. \label{c}
\end{align}
Subtracting Eq.~\eqref{a} from Eq.~\eqref{b},
\begin{equation*}
    (\lambda_1-\lambda_2)\left[3(\lambda_1+\lambda_2) - 2\beta\right] = 0.
\end{equation*}
This factorisation yields two distinct cases, creating a bifurcation point. When $\lambda_1 =\lambda_2$, the constraints give $\lambda_1=\lambda_2=a$, $\lambda_3=-2a$, and $6a^2=-2\bar{Q}_S$, so $a=\pm\sqrt{-\bar{Q}_S/3}$. The objective evaluates to
\begin{equation*}
    F = 2a^3 - 8a^3 + 2Q_\Omega a = 2a(Q_\Omega + Q_S).
\end{equation*}
Taking $a = \sqrt{-Q_S/3}$ to maximise $F$ gives
\begin{equation*}
F = \frac{2}{\sqrt{3}}(-\bar{Q}_S)^{1/2}(\bar{Q}_\Omega + \bar{Q}_S).    
\end{equation*}
This is the axisymmetric strain configuration with vorticity aligned along the symmetry axis, corresponding physically to a vortex tube.

\noindent When $\lambda_1 \neq\lambda_2$, we still have $\lambda_1+\lambda_2 = -\lambda_3$ by necessity, so $2\beta = 3(\lambda_1+\lambda_2) = -3\lambda_3$. Substituting this into Eq.~\eqref{c} allows us to eliminate the multiplier $\beta$,
\begin{equation*}
3\lambda_3^2 - \bar{Q}_\Omega - \alpha + 3\lambda_3^2 = 0 \quad\Rightarrow\quad \alpha = 6\lambda_3^2 - \bar{Q}_\Omega. 
\end{equation*}
Summing Eq.~\eqref{a} and Eq.~\eqref{b}, using $\lambda_1^2+\lambda_2^2 = -2\bar{Q}_S -\lambda_3^2$ and $\lambda_1+\lambda_2=-\lambda_3$ gives 
\begin{equation*}
3(-2\bar{Q}_S-\lambda_3^2) - 2(6\lambda_3^2-\bar{Q}_\Omega) + 3\lambda_3^2 = 0. 
\end{equation*}
\begin{equation*}
-6\bar{Q}_S - 18\lambda_3^2 + 2\bar{Q}_\Omega = 0 \quad\Rightarrow\quad\lambda_3^2 = \frac{\bar{Q}_\Omega - 3\bar{Q}_S}{9}.
\end{equation*}
For $\lambda_1\neq\lambda_2$ to be real and distinct, we require $(\lambda_1-\lambda_2)^2 > 0$, so 
\begin{equation*}
(\lambda_1-\lambda_2)^2 = (\lambda_1+\lambda_2)^2 - 4\lambda_1\lambda_2 =\lambda_3^2 - 4(\lambda_3^2+\bar{Q}_S) = -3\lambda_3^2 - 4\bar{Q}_S.    
\end{equation*}
Substituting the expression for $\lambda_3^2$ from above gives
\begin{equation*}
    (\lambda_1-\lambda_2)^2 = -\frac{\bar{Q}_\Omega}{3} - 3\bar{Q}_S  > 0.
\end{equation*}
This requires $\bar{Q}_\Omega < 9|\bar{Q}_S|$, so this case is only physically realisable in the strain-dominated regime. When $\bar{Q}_\Omega > 9|\bar{Q}_S|$, $\lambda_1$ and $\lambda_2$ become complex and this case is unphysical since it would violate the physical restrictions on the strain-rate tensor. Using $F =\lambda_3 (3\lambda_3^2+3\bar{Q}_S-\bar{Q}_\Omega)$ and substituting $3\lambda_3^2 = (\bar{Q}_\Omega-3\bar{Q}_S)/3$, the objective becomes 
\begin{equation*}
    F =\lambda_3\left(\frac{\bar{Q}_\Omega-3\bar{Q}_S}{3}+3\bar{Q}_S-\bar{Q}_\Omega\right) = -\frac{2}{3}\lambda_3(\bar{Q}_\Omega-3\bar{Q}_S).
\end{equation*}
Taking $\lambda_3 = -\frac{1}{3}\sqrt{\bar{Q}_\Omega-3\bar{Q}_S}$ (negative to maximise), the objective becomes 
\begin{equation*}
F = \frac{2}{9}(\bar{Q}_\Omega - 3\bar{Q}_S)^{3/2}    
\end{equation*}
This case has no special symmetry between $\lambda_1$ and $\lambda_2$, representing a fully triaxial strain configuration with vorticity not aligned with any principal strain axis. It exists only when $\bar{Q}_\Omega < 9|\bar{Q}_S|$ as shown above. The crossover at $\bar{Q}_\Omega = 9|\bar{Q}_S|$ marks the transition between a triaxial strain-dominated configuration and an axisymmetric vortex-tube configuration as the tightest constraint on the energy flux. We note that the triangle inequality for the local inertial flux would give 
\begin{equation*}
   |\Pi^{I, \ell}_{\mathrm{L}}|\le \frac{2}{\sqrt{3}}\ell^2(-\bar{Q}_S)^{1/2}(\bar{Q}_\Omega - \bar{Q}_S). 
\end{equation*}
It is straightforward to show that both of the bounds above are strictly tighter than this triangle inequality.

\bibliography{apssamp} 

\end{document}